\documentclass[pdflatex,sn-mathphys-num]{sn-jnl}

\usepackage{caption}
\usepackage{graphicx}%
\usepackage{multirow}%
\usepackage{amsmath,amssymb,amsfonts}%
\usepackage{amsthm}%
\usepackage{mathrsfs}%
\usepackage[title]{appendix}%
\usepackage{xcolor}%
\usepackage{textcomp}%
\usepackage{manyfoot}%
\usepackage{booktabs}%
\usepackage{algorithm}%
\usepackage{algorithmicx}%
\usepackage{algpseudocode}%
\usepackage{listings}%
\usepackage{csquotes}
\usepackage{longtable}
\usepackage{tabularray}
\usepackage{booktabs,makecell,graphicx}
\usepackage{tabularx}
\usepackage{comment}
\usepackage{lscape} 
\usepackage{amsmath, amssymb}
\usepackage{array}
\usepackage{float}     
\usepackage{geometry}
\usepackage{soul}
\usepackage{xcolor}
\usepackage{lipsum}
\theoremstyle{thmstyleone}%
\theoremstyle{thmstyletwo}%

\theoremstyle{thmstylethree}%
\newtheorem{definition}{Definition}%

\begin{document}

\title[Article Title]{A Quantitative Security Analysis of S-boxes in the NIST Lightweight Cryptography Finalists}


\author[1] {\fnm{Mahnoor} \sur{Naseer}}\email{mahnoor.naseer@uclouvain.be}

\author[2] {\fnm{Sundas} \sur{Tariq}}\email{sundas.tariq@kuleuven.be}

\author*[3] {\fnm{Naveed} \sur{Riaz}}\email{naveed.riaz@seecs.edu.pk}

\author[3] {\fnm{Naveed} \sur{Ahmed}}\email{naveed.ahmad@seecs.edu.pk}

\author[4] {\fnm{Shah} \sur{Fahd}}\email{sfahd.phd.ismcs@student.mcs.pk}

\author[5] {\fnm{Mureed} \sur{Hussain}}\email{hmureed@yahoo.com}

\author[6] {\fnm{Sajid} \sur{Ali Khan}}\email{sajid.ali@fui.edu.pk}

\affil[1] {\orgdiv{Crypto Group, ICTEAM Institute}, \orgname{UCLouvain}, \city{Louvain-la-Neuve}, \country{Belgium}}

\affil[2] {\orgdiv{3MI Labs and COSIC}, \orgname{KU Leuven}, \city{Leuven},  \country{Belgium}}
 
\affil[3] {\orgdiv{SEECS}, \orgname{National University of Sciences and Technology}, \city{Islamabad}, \country{Pakistan}}

\affil[4] {\orgdiv{MCS}, \orgname{National University of Sciences and Technology}, \city{Islamabad}, \country{Pakistan}}

\affil[5] {\orgdiv{National Center for Cyber Security (NCCS)}, \orgname{Air University}, \city{Islamabad}, \country{Pakistan}}

\affil[6] {\orgdiv{Department of Software Engineering}, \orgname{Foundation University}, \city{Islamabad}, \country{Pakistan}}


\abstract{Lightweight cryptography was primarily inspired by the design criteria of symmetric cryptography. It plays a vital role in ensuring the security, privacy, and reliability of microelectronic devices without compromising the overall functionality and efficiency. However, the increasingly platform specific design requirements prompted the development of a standard lightweight algorithm. In 2017, NIST put forward security requirements for a standard lightweight scheme \textemdash\ security strength of at least 112 bits against known cryptanalysis attacks, mitigation against side channel and fault injection attacks, and implementation efficiency. After three rounds of review, ASCON was crowned as the winner of the competition. Evaluating the individual components used in any cryptographic algorithm is an important step in the verification of security claims. A fundamental component used to ensure Shannon's property of confusion in cryptographic primitives is an S-box. Hence, the quality of an S-box is a significant contributing factor in the security strength of a cipher. In this paper, we evaluate the S-boxes of $\textbf{6}$ NIST LWC competition finalists based on well-known cryptographic properties, and comment on how the results reflect upon NIST security requirements. Our findings have revealed that these S-boxes do not comply with the basic notions of avalanche, making it vulnerable to high-order sophisticated cryptanalysis.}


\maketitle

\section{Introduction}\label{sec1}

Application specific computing has become an emerging trend in modern information technology due to the wide usability of smart devices. The highly evolving yet resource constrained environment enumerates to countless security concerns. It is difficult to use traditional cryptographic primitives on these platforms while striking an optimal trade-off between compact implementation and security strength \mbox{\cite{RezaMehran}}. In order to effectively resolve this issue, NIST laid the foundation for lightweight cryptography standardization in $2017$ \cite{bassham2018submission}.\\
Lightweight cryptography aims to provide security for Internet of Things (IoT) devices, namely sensors, RFID tags, medical implants, smart cards, etc. Design choices such as small block and key size, simple round structure and key schedule, and nominal implementation make lightweight ciphers the right choice to ensure a sufficient level of security using less power, computing, and memory resources.\\
As a result of NIST's call, a total of $57$ proposals were submitted by research teams across the globe. After two rounds of the rigorous evaluation process, as described in NIST IR $8369$ \cite{turan2021status}, $10$ candidates were shortlisted as finalists of the competition (see Table \ref{table:1} \cite{finalist}).
In the third and final round, candidates were evaluated on the basis of claims made in terms of  security strength against known attacks ($112$ bits or more); performance in both hardware and software environments; resilience against side channel and fault injection attacks; and intellectual property rights \cite{turan2023status}.
\\
In February $2023$, NIST declared the ASCON family as the new lightweight cryptography standard \cite{result}.
The most important evaluation criteria for cryptographic algorithms is measuring their security strength against known attacks such as differential and linear cryptanalysis (and their variants), as well as implementation based attacks. Designers and cryptanalysts across the globe tried their level best to suffice this requirement. As a result, the competition finalists received a noticeable amount of third-party security analysis \cite{cihangir1, li2017cryptanalysis, dobaruyng, Duarte-Sanchez2021, Erlacher, ElHirch_Mella, Leander, 9474092, Christoph2015, Kannwischer, udvarhelyi2021security, yaxin, Meichun,Eskandri, ji2020improved, Sun_Wang_Wang_2021, cryptoeprint:2018/390, cryptoeprint:2021/737, cryptoeprint:2020/1405, 6855573, Jana, dobraunigkey, Akiko,JGuo,cuitinting,Jean, zhou2021interpolation, Priyanka, meraneh2022blind, Beyne, Gorjan, Tolba, rom1, rom2, Hosein, Shi, Qin_Dong, Bijwe}, a few highlighted in Table \ref{table:2}:-

\renewcommand{\arraystretch}{1.5}
\begin{table*}[h]
\centering
\caption{Classification of NIST LWC Finalists based on Presence/Absence of S-box}
\label{table:1}
\begin{tabular}{ll} 
\hline
With S-box & Without S-box                                              \\ \hline
\begin{tabular}[c]{@{}l@{}}ASCON, ISAP, GIFT-COFB, \\Photon-Beetle, Elephant, Romulus~\end{tabular} & 
\begin{tabular}[c]{@{}l@{}}Grain-$128$-AEAD, Sparkle, \\Xoodyak, TinyJAMBU\end{tabular} \\
\hline
\end{tabular}
\end{table*}

\renewcommand{\arraystretch}{1.5} 
\begin{table*}[h]
\centering 
\caption{Summary of Third-Party Security Analysis on NIST LWC Finalists}
\label{table:2}
\resizebox{\textwidth}{!}{%
\begin{tabular}{l p{4cm} p{3cm} 
p{5.5cm} p{3.8cm}}
\toprule
\textbf{Cipher} & \textbf{Attack Type} & \textbf{Rounds} & \textbf{Complexity} & \textbf{Reference} \\
\midrule

\multirow{5}{*}{ASCON} 
& Differential-Linear & \parbox{3cm}{$4$ \\ $5$} & \parbox{5.5cm} {Time $= 2^{15}$, $2^{18}$ \\ Time $= 2^{33.1}$, $2^{36}$} & \parbox{3.8cm}{\cite{cihangir1}\\ \cite{dobaruyng}} \\
& \parbox{5cm}{Impossible Differential \\ Truncated Differential} 
& \parbox{3cm}{$4, 5$ \\ $4, 5$} & \parbox{5.5cm}{Data $= 2^{2}$, $2^{109}$ \\ Data $= 2^{2}$, $2^{109}$} & \cite{cihangir2} \\
& Cube Attack & \parbox{3cm}{$5$ \\ $6$} & \parbox{5.5cm}{Time $= 2^{35}$ \\ Time $= 2^{66}$} & \cite{dobaruyng} \\
& \parbox{5cm}{Key Recovery \\ (Nonce Misuse)} & \parbox{3cm}{Init ($7/12$) \\ $8,9$} & Time $= 2^{97}$, $2^{101}$, $2^{123.92}$ & \parbox{3.8cm}{\cite{li2017cryptanalysis} \\ \mbox{\cite{peng2025improved}}} \\
& Forgery Attack & Final $(4-6)$ & Time $= 2^9$, $2^{17}$ \& $2^{33}$ & \cite{li2017cryptanalysis} \\
& Round Reduction $+$ Cube & Init $(5)$ & Time $= 2^{24}$ & \cite{Duarte-Sanchez2021} \\ 
& \parbox{5cm}{CPA using Multi-bit \\Selection Function} & Full & \parbox{5cm}{$2$-bit selection function $= 7200$ traces \\ $3$-bit selection function $= 8200$ traces} & \mbox{\cite{nguyencorrelation}}\\
& Differential Fault Attack & Full & Fault queries $= 2^{10}$ & \mbox{\cite{jana2024differentialascon}}\\
& Collision Attack & $2$ & Time $= 2^{98}$ & \mbox{\cite{zhai2024improved}} \\
& Free-start Collison Attack & $3$ & Time $= 2^{10}$ & \mbox{\cite{fu2025preimage}} \\
\hline

\multirow{4}{*}{GIFT-128 / COFB}
& \parbox{5cm}{MILP-Based \\Linear Attack} & $20$ & $65$-bit key recovery & \cite{yaxin} \\
& Differential Key Recovery Attack & $23$ & Time $= 2^{120}$, Data $= 2^{86}$  & \cite{Baoyu} \\
& \parbox{5cm}{Integral Distinguisher\\ using Division Property} & $11$ & $-$ & \cite{Eskandri} \\
& Differential Attack & $26$ & Time $= 2^{123.245}$  & \cite{ji2020improved} \\
\hline

\multirow{3}{*}{Photon-Beetle} 
& Key Recovery Irregularity & Full & Data $= 2^{50}$ bytes & \cite{dobraunigkey} \\
& Differential Fault Attack & Full & Time $= 2^{16}$, Fault queries $= 2^{37.15}$ & \mbox{\cite{jana2024differential}} \\ 
& Committing Attack & Initial State & No queries & \mbox{\cite{Krämer_Struck_Weishäupl_2024}} \\
& Pre-image Attack & $3.5$ & Time = $2^{112}$, Data $= 2^{65}$ & \mbox{\cite{dong2024generic}} \\
\hline

\multirow{3}{*}{Elephant}
& Interpolation Attack & $8$ & Time $= 2^{70}$ & \cite{zhou2021interpolation} \\
& Fault Analysis (Dumbo) & Full & $85 - 250$ ciphertexts & \cite{Priyanka} \\
& Blind SCA on LFSR & Full & $<2$ days (practical) & \cite{meraneh2022blind} \\ 
\hline

\multirow{3}{*}{Romulus} 
& Impossible Differential & $22$ & Time $= 2^{373.48}$ , Data $= 2^{92.22}$ & \cite{Tolba} \\
& Matching / Authenticity Attack & Full & $-$ & \cite{rom1} \\
& Meet-in-the-Middle (MITM) & $23$ & Time $= 2^{376}$ , Data $= 2^{104}$ & \cite{rom2} \\
& Impossible Boomerang & \parbox{3cm}{$27$ \\ $28$} & \parbox{5.5cm}{Time $= 2^{337}$, Data $= 2^{131.3}$ \\ Time $= 2^{382.8}$, Data $= 2^{130.26}$} & \mbox{\cite{Zhang_Wang_Tang_2024}}  \\ 
& Differential MITM & 25 & Time $= 2^{366}$, Data $= 2^{122.3}$ & \mbox{\cite{10.1007/978-3-031-58716-0_10}}  \\ 
& \parbox{5cm}{Impossible Differential \\ MITM} & \parbox{3cm}{$21$ \\ $22$ \\ $23$ \\ $27$\\ $28$} & \parbox{5.5cm}{Time $= 2^{344.33}$, Data $= 2^{122.89}$ \\Time $= 2^{378.22}$, Data $= 2^{121.50}$ \\ Time $= 2^{378.22}$, Data $= 2^{121.50}$ \\ Time = $2^{361.06}$, Data = $2^{124.99}$ \\ Time $= 2^{378.22}$, Data $= 2^{129.50}$} & \mbox{\cite{song2025generalized}} \\ 
\bottomrule
    \end{tabular}
    }
\end{table*}

Shannon's property confusion in any symmetric cipher is mainly dependent on the non-linear component used. The most commonly used component to achieve non-linearity in symmetric cryptography is a substitution box (S-box).
Analysing the cryptographic profile of an S-box plays a vital role in determining the overall strength of a cryptosystem. The literature reports nearly $24$ cryptographic properties
\cite{perrin2013properties,heys2002tutorial,ishfaq2018matlab, seitkulov2021,nitaj2020new,abeer,zahid2021dynamic,evertse1987linear,nyberg2023modifications,boura2018boomerang,tezcan2014differential, bar2019dlct,kim2018improved,canteaut2016lecture, guilley2004differential,prouff2005dpa} that can be used as a tool to estimate the quality of an S box. 
Some of the attacks mentioned earlier exploit specific S-box properties in order to devise practical attack characteristics for a particular cipher. This paper aims to provide a complete picture of the cryptographic properties of S-boxes, their relevance to known attacks, and related security limits \footnote{Current paper is an extended version of our previous paper \cite{9990069}}. Our primary goal is to help researchers better understand and evaluate the security profile of S-boxes in a principle manner.  
 
Section \ref{sec2}  of this paper elaborates the specifications of six finalists of NIST LWC. Section \ref{sec3} discusses the fundamental properties of S boxes and classifies them according to the relevant attack category. Secondly, the upper and lower bounds of these properties are highlighted along with their behavior under affine equivalence. In Section \ref{sec4}, the results of the S-box analysis are discussed in detail. In Section \ref{sec5}, we generate a new $4-$bit S-box with optimal cryptographic properties using the PEIGEN S-box generation tool \cite{peigen}, and draw a comparison of the properties of this S-box with the LWC finalists having $4-$bit S-boxes. Section \ref{sec6} discusses and highlights the scope and limitations of this paper. Finally, we conclude our findings in Section \ref{sec7}.

\section{Overview of Finalists} \label{sec2}
This section provides a brief introduction to each of the $6$ finalists from NIST lightweight standardization that make use of S-box as one of the design components.

\subsection{ASCON}
It is a cryptographic cipher suite presented by Maria \textit{et al}. \cite{dobraunig2021ascon}.  ASCON comprises of three authenticated encryption and associated data (AEAD) variants: $128$, $128$ ($a$ and $pq$), and two hashing algorithms: Hash and Hash-$a$. Its main functionality comes from the use of a $320$ bit permutation, which is combined with different constants and number of rounds, depending upon the variant. In addition, to achieve non-linearity in the encryption algorithm a $5 \times 5$ affine equivalent S-box of $\chi$ mapping from Keccak permutation is used. 
The ASCON family has been selected by NIST for lightweight standardization because of its security strength and flexibility to be used in multiple applications while bearing a minimum implementation cost \cite{result}. 
\subsection{ISAP}
ISAP was proposed by S. Mangard \textit{et al}. \cite{dobraunig2020isap}, it is a nonce based AEAD cryptographic scheme designed to resist against implementation based attacks at an optimized implementation cost. ISAP uses $400$ bit Keccak and $320$ bit ASCON permutations in its design alongside $5 \times 5$  bit ASCON S-box as a non-linear component. 

\subsection{GIFT-COFB}
Proposed by A. Chakraborti. \textit{et al}.\cite{cryptoeprint:2020:738}. GIFT-COFB is an encryption scheme that uses combined feedback mode with GIFT-$128$. GIFT-COFB employs $40$ rounds of encryption where the input is: an arbitrary length associated data, an input message of variable length, a nonce of $128$ bits, and a $128$ bit encryption key. Whereas the output is a $128$ bit tag and ciphertext of same length as the input message. A $4 \times 4$ bit S-box is used as the confusion component.

\subsection{Photon-Beetle}
Photon-Beetle is an AEAD based hashing and encryption scheme presented by J. Guo \textit{et al}. \cite{bao2019photon}. The encryption routine comprises of $12$ rounds consisting of $4$ separate stages i.e. XOR constant, substitute cells, shift rows and serial mix columns. Photon-Beetle (AEAD) uses the $4$ × $4$ bit S-box of Present block cipher in order to achieve non-linearity in its design.
\subsection{Elephant} 
C. Dobarunig \textit{ et a}l. \cite{mennink2021elephant}, proposed Elephant as a permutation based AEAD algorithm. It has $3$ variants based on Spongent and Keccak permutations: Delirium, Jumbo and Dumbo. These encryption variants take $128$ bit key and $96$ bit nonce, however the block sizes are $160$, $176$, $200$. A $4 \times 4$ S-box as the non-linear part. In the $80$ rounds of encryption there are $160$ differentially active S-boxes. 

\subsection{Romulus}
T. Iwata. \textit{et al}. \cite{guodesigners}, proposed Romulus as a nonce based AEAD encryption and hashing scheme that uses Skinny block cipher as underlying encryption algorithm.  It operates in $4$ modes of encryption and hashing: nonce misuse based Romulus-M, nonce based Romulus-N, leakage resistant based Romulus-T and hashing based Romulus-H. The encryption algorithm takes $128$ bits of key and nonce, and runs $40$ rounds of encryption. Skinny-$128$ tweakable block cipher \cite{Skinny}  uses $8 \times 8$ S-box to achieve non-linearity in its design.


\renewcommand{\arraystretch}{1.5}

\begin{table*}[h]
  \centering
  \caption{Classification of S-box properties as per relevance to existing cryptanalysis techniques}
  \label{table:3}
    \begin{tabular}{@{}lll@{}}
      \toprule
      \textbf{S\#} & \textbf{Criteria}               & \textbf{Properties}  \\
      \midrule
      1  & Generic                       & \begin{minipage}[t]{0.7\textwidth}
                                         Balancedness; Permutation; Order of Permutation;\\
                                         Fixed Points (FP); Opposite Fixed Points (OFP);\\
                                         Bit Independence Criteria (BIC);\\
                                         Strict Avalanche Criteria (SAC);\\
                                         Auto Correlation Table (ACT) \cite{joscubero};\\
                                         Absolute Indicator (AI) \cite{khadem2021construction};\\
                                         Sum of Square Indicators (SSI) \cite{khadem2021construction}
                                       \end{minipage} \\
\hline
      2  & Linear Cryptanalysis         & \begin{minipage}[t]{0.7\textwidth}
                                         Nonlinearity (NL) \cite{nitaj2020new,abeer};\\
                                         Linear Approximation Table (LAT) \cite{zahid2021dynamic,10.1007/3-540-60590-8_10};\\
                                         Linear Approximation Probability (LAP);\\
                                         Linear Branch Number (LBN);\\
                                         Linear Structures (LS) \cite{evertse1987linear,nyberg2023modifications};\\
                                         Correlation Immunity (CI) \cite{bakunina2022synthesis}
                                       \end{minipage} \\
\hline
      3  & Differential Cryptanalysis   & \begin{minipage}[t]{0.7\textwidth}
                                         Differential Distribution Table (DDT) \cite{zahid2021dynamic};\\
                                         Differential Uniformity (DU) \cite{nitaj2020new,abeer};\\
                                         Differential Branch Number (DBN);\\
                                         Propagation Characteristics (PC) \cite{alvarez2012cryptographic};\\
                                         Undisturbed Bits \cite{tezcan2014differential}
                                       \end{minipage} \\
                                       
\hline
      4  & Boomerang Cryptanalysis      & \begin{minipage}[t]{0.7\textwidth}
                                         Boomerang Connectivity Table (BCT);\\
                                         Boomerang Uniformity (BU) \cite{boura2018boomerang}
                                       \end{minipage} \\
\hline
      5  &  Differential–Linear Cryptanalysis  & \begin{minipage}[t]{0.7\textwidth}
                                         Linear Structures (LS) \cite{evertse1987linear,nyberg2023modifications};\\
                                         Differential Linear Connectivity Table (DLCT) \cite{bar2019dlct};\\
                                         Differential Linear Uniformity (DLU) \cite{kim2018improved}
                                       \end{minipage} \\
\hline
      6  & Algebraic Cryptanalysis      & \begin{minipage}[t]{0.7\textwidth}
                                         Algebraic Degree (AD) \cite{canteaut2016lecture};\\
                                         Interpolation Polynomial (IP)
                                       \end{minipage} \\
\hline
      7  & Side-Channel Analysis        & \begin{minipage}[t]{0.7\textwidth}
                                         DPA-SNR \cite{guilley2004differential};\\
                                         Transparency Order (TO) \cite{prouff2005dpa}
                                       \end{minipage} \\
      \bottomrule
    \end{tabular}%
\end{table*}


\renewcommand{\arraystretch}{1.5}
\begin{table*}[h]
  \centering
  \caption{Theoretical Bounds on the Cryptographic Properties of S-boxes}
  \label{table:4}
  \resizebox{\textwidth}{!}{%
    \begin{tabular}{|l|l|c|c|c|c|c|}
      \toprule
      \textbf{S\#} & \textbf{Properties}
        & \multicolumn{1}{c|}{\textbf{Bounds}}
        & \multicolumn{1}{c|}{\textbf{\begin{tabular}[c]{@{}c@{}}For\\4×4\end{tabular}}}
        & \multicolumn{1}{c|}{\textbf{\begin{tabular}[c]{@{}c@{}}For\\5×5\end{tabular}}}
        & \multicolumn{1}{c|}{\textbf{\begin{tabular}[c]{@{}c@{}}For\\8×8\end{tabular}}}
        & \multicolumn{1}{c|}{\textbf{\begin{tabular}[c]{@{}c@{}}Ideal\\Value\end{tabular}}} \\
      \midrule
1  & $\mathsf{OP}$ & $1 \le \mathsf{OP} \le 2^n$  & $1 \le \mathsf{OP} \le 16$ & $1 \le \mathsf{OP} \le 32$ & $1 \le \mathsf{OP} \le 256$ & Close to UB \\  \hline
2  & $\mathsf{FP}$ & $0 \le \mathsf{FP} \le 2^n$ & $0 \le \mathsf{FP} \le 16$ & $0 \le \mathsf{FP} \le 32$  & $0 \le \mathsf{FP} \le 256$  & Close to LB \\ \hline
3  & $\mathsf{OFP}$ & $0 \le \mathsf{OFP} \le 2^n$ & $0 \le \mathsf{OFP} \le 16$  & $0 \le \mathsf{OFP} \le 32$  & $0 \le \mathsf{OFP} \le 256$ & Close to LB \\ \hline
4  & $\mathsf{BIC}$  & $0 \le \mathsf{BIC} \le 1$ & $0 \le \mathsf{BIC} \le 1$  & $0 \le \mathsf{BIC} \le 1$   & $0 \le \mathsf{BIC} \le 1$   & $0$ \\ \hline
5  & $\mathsf{SAC}$  & $0 \le \mathsf{SAC} \le 1$ & $0 \le \mathsf{SAC} \le 1$  & $0 \le \mathsf{SAC} \le 1$   & $0 \le \mathsf{SAC} \le 1$   & $0.5$ \\ \hline
6  & $\barwedge_S$   & $\sqrt{\tfrac{2^{2n}}{2^n-1}} \le \barwedge_S \le 2^n$ & $0 < \barwedge_S \le 16$  & $0 < \barwedge_S \le 32$  & $0 < \barwedge_S \le 256$  & Close to LB \\ \hline
7  & $\doublebarwedge_S$ & $2^{2n} \le \doublebarwedge_S \le 2^{3n+m}$ & $256 \le \doublebarwedge_S \le 65536$ & $1024 \le \doublebarwedge_S \le 1048576$ & $65536 \le \doublebarwedge_S \le 4294967296$ & Close to LB \\
      \hline
8  & $\mathsf{LAT}$ & $0 \le \mathsf{LAT} \le 2^n-1$ & $0 \le \mathsf{LAT} \le 15$  & $0 \le \mathsf{LAT} \le 31$  & $0 \le \mathsf{LAT} \le 255$ & - \\ \hline
9  & $\xi$ & $0 \le \xi \le \tfrac12$ & $0 \le \xi \le \tfrac12$    & $0 \le \xi \le \tfrac12$    & $0 \le \xi \le \tfrac12$    & Close to LB \\ \hline
10 & $\mathsf{NL}$ & \begin{tabular}[c]{@{}l@{}} For $n$ even: $0\le\mathsf{NL}<2^{n-1}-2^{\tfrac n2-1}$\\ For $n$ odd:  $0\le\mathsf{NL}<2^{n-1}-2^{\tfrac{n-1}2}$\end{tabular} & $0\le\mathsf{NL}<6$ & $0\le\mathsf{NL}<12$  & $0\le\mathsf{NL}<120$ & Close to UB \\
      \hline
11  & $\mathsf{BN_L}$ & $2 \le \mathsf{BN_L} \le n-1$ & $2 \le \mathsf{BN_L} \le 3$  & $2 \le \mathsf{BN_L} \le 4$  & $2 \le \mathsf{BN_L} \le 7$  & Close to UB \\ \hline
12  & $\mathsf{LS}$   & $0 \le \mathsf{LS} \le 2^{n+m}$ & $0 \le \mathsf{LS} \le 256$  & $0 \le \mathsf{LS} \le 1024$ & $0 \le \mathsf{LS} \le 65536$& Close to LB \\ \hline
13  & $\mathsf{CI}$   & $0 \le \mathsf{CI} \le n$ & $0 \le \mathsf{CI} \le 4$ & $0 \le \mathsf{CI} \le 5$ & $0 \le \mathsf{CI} \le 8$ & Close to LB \\ \hline
14  & $\mathsf{DDT}$  & $0 \le \mathsf{DDT} \le 2^n-1$ & $0 \le \mathsf{DDT} \le 15$  & $0 \le \mathsf{DDT} \le 31$  & $0 \le \mathsf{DDT} \le 255$ & - \\ \hline
15  & $\mathsf{DU}$   & $2 \le \mathsf{DU} \le 2^n$ & $2 \le \mathsf{DU} \le 16$   & $2 \le \mathsf{DU} \le 32$   & $2 \le \mathsf{DU} \le 256$  & Close to LB \\ \hline
16  & $\mathsf{BN_D}$ & $2 \le \mathsf{BN_D} \le \lceil\tfrac{2n}{3}\rceil$ & $2 \le \mathsf{BN_D} \le 3$  & $2 \le \mathsf{BN_D} \le 4$  & $2 \le \mathsf{BN_D} \le 6$   & Close to UB \\ \hline
17  & $\mathsf{PC}$   & $0 \le \mathsf{PC} \le n$ & $0 \le \mathsf{PC} \le 4$  & $0 \le \mathsf{PC} \le 5$    & $0 \le \mathsf{PC} \le 8$    & Close to UB \\ \hline
18  & $\mathsf{UDB}$  & $\mathsf{UDB}=0$ & $\mathsf{UDB}=0$ & $\mathsf{UDB}=0$  & $\mathsf{UDB}=0$    & $0$ \\ \hline
19  & $\beta\mathsf{U}$  & $2 \le \beta\mathsf{U} \le 2^n$ & $2 \le \beta\mathsf{U} \le 16$& $2 \le \beta\mathsf{U} \le 32$& $2 \le \beta\mathsf{U} \le 256$& Close to LB \\
      \hline
20  & $\mathsf{DLU}$ & $2^{\tfrac n2-1}<\mathsf{DLU}\le2^{n-1}$ & $2<\mathsf{DLU}\le8$  & $4<\mathsf{DLU}\le16$   & $8<\mathsf{DLU}\le128$  & Close to LB \\ \hline
21  & $\mathsf{AD}$  & $1 \le \mathsf{AD} \le n-1$ & $1 \le \mathsf{AD} \le 3$    & $1 \le \mathsf{AD} \le 4$  & $1 \le \mathsf{AD} \le 7$    & Close to UB \\ \hline
22 & $\mathsf{DPA_{SNR}}$   & \begin{tabular}[c]{@{}l@{}} For balanced S-box: \\ $1 \le \mathsf{DPA_{SNR}} \le 2^{\tfrac n2}$\end{tabular}  & $0 \le \mathsf{DPA_{SNR}} \le 4$ & $0 \le \mathsf{DPA_{SNR}} < 6$ & $0 \le \mathsf{DPA_{SNR}} \le 16$ & Close to LB \\  \hline
23 & $\mathsf{TO}$& $0 \le \mathsf{TO} \le m$ & $0 \le \mathsf{TO} \le 4$  & $0 \le \mathsf{TO} \le 5$ & $0 \le \mathsf{TO} \le 8$ & Close to LB \\
    \bottomrule
    \end{tabular}
  }
\end{table*}

\renewcommand{\arraystretch}{1.5}
\begin{table*}[h]
\centering
\caption{S-box Properties under Affine Equivalence}
\label{table:5}%
\begin{tabular}{@{}lcc@{}}
\toprule
\textbf{Property} & \textbf{Affine Variant} & \textbf{Affine Invariant} \\
\midrule
    Bijectivity                &                     & \checkmark \\
    Balancedness               &                     & \checkmark \\
    Permutation                & \checkmark          &             \\
    $\mathsf{OP}$              & \checkmark          &             \\
    $\mathsf{FP}$              & \checkmark          &             \\
    $\mathsf{OFP}$             & \checkmark          &             \\
    $\mathsf{BIC}$             & \checkmark          &             \\
    $\mathsf{SAC}$             & \checkmark          &             \\
    $\mathsf{ACT}$             &                     & \checkmark \\
    $\barwedge$                &                     & \checkmark \\
    $\doublebarwedge$          &                     & \checkmark \\
    $\mathsf{NL}$              &                     & \checkmark \\
    $\mathsf{LAT}$             &                     & \checkmark \\
    $\mathsf{BN\_L}$           & \checkmark          &             \\
    $\mathsf{LS}$              & \checkmark          &             \\
    $\mathsf{CI}$              & \checkmark          &             \\
    $\mathsf{DDT}$             &                     & \checkmark \\
    $\mathsf{DU}$              &                     & \checkmark \\
    $\mathsf{BN\_D}$           & \checkmark          &             \\
    $\mathsf{PC}$              &                     & \checkmark \\
    $\mathsf{UDB}$             &                     & \checkmark \\
    $\mathsf{BCT}$             &                     & \checkmark \\
    $\beta\mathsf{U}$          &                     & \checkmark \\
    $\mathsf{DLCT}$            &                     & \checkmark \\
    $\mathsf{DLU}$             &                     & \checkmark \\
    $\mathsf{AD}$              &                     & \checkmark \\
    $\mathsf{IP}$              &                     & \checkmark \\
    $\mathsf{DPA\text{-}SNR}$  & \checkmark          &             \\
    $\mathsf{TO}$              & \checkmark          &             \\
\botrule
\end{tabular}
\end{table*}

\begin{table*}[h]
  \centering
  \caption{Results of S-box Analysis -- $8\times 8$ and $5 \times5$ bit case}
  \label{table:6}
  \scalebox{0.85}{
    \begin{tabular}{!{\color{black}\vrule}l
    !{\color{black}\vrule}l
    !{\color{black}\vrule}c
    !{\color{black}\vrule}c
    !{\color{black}\vrule}c
    !{\color{black}\vrule}c
    !{\color{black}\vrule}c}
      \toprule
      {\textbf{S\#}} 
        & {\textbf{Properties}} 
        & \multicolumn{1}{c!{\color{black}\vrule}}{\begin{tabular}[c]{@{}c@{}}\textbf{Romulus}\\\textbf{$8\times8$ Skinny-128}\end{tabular}}
        & \multicolumn{1}{c!{\color{black}\vrule}}{\begin{tabular}[c]{@{}c@{}}\textbf{ASCON/ISAP}\\\textbf{$5\times5$}\end{tabular}} \\
      \midrule
1  & Bijectivity   & \checkmark  & \checkmark  \\
2  & Balancedness  & \checkmark  & \checkmark  \\
3  & Permutation   & \checkmark   & \checkmark \\
4  & $\mathsf{OP}$ & 140 & 26          \\
5  & $\mathsf{FP}$ & 1   & 0           \\
6  & $\mathsf{OFP}$& 0   & 0         \\
7  & $\mathsf{BIC}$& 1.000000  & 1.000000     \\
8  & $\mathsf{SAC}$& 1.000000  & 1.000000   \\
9  & $\barwedge_S$ & 256       & 32          \\
10 & $\doublebarwedge_S$  & 4194304 & 8192  \\
11 & $\xi$                & 0.25000 & 0.25000     \\
12 & $\mathsf{NL}$        & 64      & 8      \\
13 & $\mathsf{BN_L}$      & 2       & 3        \\
14 & $\mathsf{LS}$        & 601     & 91     \\
15 & $\mathsf{CI}$        & 0       & 0          \\
16 & $\mathsf{DU}$        & 64      & 8          \\
17 & $\mathsf{BN_D}$      & 2       & 3           \\
18 & $\mathsf{PC}$        & 0       & 0           \\
19 & $\mathsf{UDB}$       & 258     & 35          \\
20 & $\beta\mathsf{U}$    & 256     & 16     \\
21 & $\mathsf{DLU}$       & 128    & 16    \\
22 & $\mathsf{AD}$        & 6       & 2     \\
23 & $\mathsf{DPA\mbox{-}SNR}$      & 6.312455 & 3.015113  \\
24 & $\mathsf{TO}$        & 7.174510 & 4.258065     \\ 
\bottomrule
\end{tabular}%
  }
\end{table*}


\begin{table*}[h]
  \centering
  \caption{Results of S-box Analysis -- $4\times 4$ bit case}
  \label{table:7}
  \scalebox{0.85}{
    \begin{tabular}{!{\color{black}\vrule}l
    !{\color{black}\vrule}l
    !{\color{black}\vrule}c
    !{\color{black}\vrule}c
    !{\color{black}\vrule}c
    !{\color{black}\vrule}c
    !{\color{black}\vrule}c}
      \toprule
      {\textbf{S\#}} 
        & {\textbf{Properties}} 
          & \multicolumn{1}{c!{\color{black}\vrule}}{\begin{tabular}[c]{@{}c@{}}\textbf{New Sbox }\\\textbf{$4 \times 4$}\end{tabular}}
        & \multicolumn{1}{c!{\color{black}\vrule}}{\begin{tabular}[c]{@{}c@{}}\textbf{Elephant}\\\textbf{$4\times4$}\end{tabular}}
        & \multicolumn{1}{c!{\color{black}\vrule}}{\begin{tabular}[c]{@{}c@{}}\textbf{GIFT-COFB}\\\textbf{$4\times4$ GIFT}\end{tabular}}
        & \multicolumn{1}{c!{\color{black}\vrule}}{\begin{tabular}[c]{@{}c@{}}\textbf{Photon-Beetle}\\\textbf{$4\times4$ Present}\end{tabular}} \\ 
      \midrule
1  & Bijectivity  & \checkmark & \checkmark & \checkmark & \checkmark \\
2  & Balancedness  & \checkmark & \checkmark & \checkmark & \checkmark \\
3  & Permutation   & \checkmark & \checkmark & \checkmark & \checkmark \\
4  & $\mathsf{OP}$ &   7  & 13   & 9   & 7       \\
5  & $\mathsf{FP}$ &   3   & 0    & 0    & 0          \\
6  & $\mathsf{OFP}$&   1    & 1    & 1    & 1 \\
7  & $\mathsf{BIC}$&  0.577350    & 1.000000   & 1.000000   & 1.000000   \\
8  & $\mathsf{SAC}$&  0.7500    & 1.000000   & 1.000000   & 1.000000   \\
9  & $\barwedge_S$ &   8   & 16  & 16  & 16       \\
10 & $\doublebarwedge_S$ &  640   & 1024 & 1024  & 1024 \\
11 & $\xi$&  0.25000   & 0.25000    & 0.25000    & 0.25000    \\
12 & $\mathsf{NL}$ &   4   & 4  & 4   & 4          \\
13 & $\mathsf{BN_L}$ & 2   & 2  & 2   & 2          \\
14 & $\mathsf{LS}$ &   0   & 9  & 9   & 9          \\
15 & $\mathsf{CI}$ &   0   & 0  & 0   & 0          \\
16 & $\mathsf{DU}$ &   4   & 4  & 6   & 4          \\
17 & $\mathsf{BN_D}$ &  2  & 3  & 2   & 3          \\
18 & $\mathsf{PC}$ &  0    & 0  & 0   & 0          \\
19 & $\mathsf{UDB}$&  0    & 3  & 6   & 6          \\
20 & $\beta\mathsf{U}$ &   10  & 16 & 16  & 16        \\
21 & $\mathsf{DLU}$ &   4  & 8& 8& 8 \\
22 & $\mathsf{AD}$ &     3    & 3  & 3   & 3          \\
23 & $\mathsf{DPA\mbox{-}SNR}$ &   2.578634     & 2.398501 & 2.398501  & 2.128608 \\
24 & $\mathsf{TO}$ &     3.400000      & 3.266667  & 3.466667   & 3.533333   \\ 
      \bottomrule
    \end{tabular}%
  }
\end{table*}

\section{Cryptographic Properties of S-boxes and their Classification}\label{sec3}
The strength of cryptographic primitives is best analysed with the help of well-known cryptanalysis techniques. Shannon's theory of confusion and diffusion plays the foundational role in the design and analysis of iterative ciphers. The aspect of confusion is introduced in symmetric ciphers using highly non-linear components such as S-boxes. Designers and cryptanalysts treat these ciphers as a white box, analysing all the underlying mathematical components in order to understand and rule out any potential attack vectors, by imposing design practices based on standard characteristics and theoretically well-defined bounds.
Owing to the importance of S-boxes in the design of cryptographic algorithms, an extensive research has been done over the years to explain the most refined design and analysis criteria that could be employed in order to secure them against potential attack vectors. These properties can be classified in a number of ways depending whether upon the types of representation in boolean algebra or the statistical and implementation based attacks that could be launched through exploitation of relevant metrics or via studying the inter-dependencies that link one property to another. 

This section briefly covers an introduction to around twenty-four cryptographic properties of S-boxes that hold significance in the current literature. Each of the properties is related to one or more crytanalysis techniques, that could exploit the non-linear part of a block cipher, see Table \ref{table:3}. In addition, Table \ref{table:4} presents the upper and lower bounds (UB \& LB) on these properties with respect to the dimension of S-box being used \cite{carlet2021boolean,sarkar2017bounds,canteaut2016lecture, song2019boomerang,canteaut2021autocorrelations,canteaut2019differential,guilley2004differential, prouff2005dpa}. We also talk about the behaviour of these properties under affine equivalence, see Table \ref{table:5}. The properties along with their respective bounds and equivalence characteristics aid designers and cryptanalysts in the selection of good S-boxes.

\subsection{Diffusion Characteristics}
Properties such as balancedness, bijectivity, order of permutation, fixed and opposite fixed points, strict avalanche and bit independence criteria, absolute and sum of square indicator are general pre-requisites that should adhere to defined bounds for any S-box in order to avoid correlations between the input and output bits and serve as a pre-requisite to ensure security against various statistical cryptanalysis attacks.

\paragraph{Balancedness}
A vectorial boolean mapping $S: F_2^n  \rightarrow F_2^m $ is balanced if it has an equal number of $1$’s and $0$’s in its coordinate and component functions. Mathematically; 
\small
\begin{equation}
\label{eq:1}
P(0) = P(1) = \frac{1}{2}= 50\%
\end{equation}
An imbalanced boolean mapping stays at a risk of achieving high linear approximation probabilities \cite{farwa2016highly}. Balancedness is invariant when under affine equivalence, which means that the number of $0$'s ad $1$'s is preserved in $S$ if an affine transformation is applied to it  \cite{sagdiccoglu2003cryptological,picek2016search,musukwa2019some}.
\paragraph{Permutation} 
A boolean mapping $S: F_2^n  \rightarrow F_2^m$ is permutation when the entire input space is utilized in such a way that one input maps only at one output. To fulfill this property it is necessary for $S$ to be a bijective mapping i.e. $n=m$. Permutation property of $S$ is variant under affine transformation \cite{nitaj2020new,abeer,cui2011improved,seghier2019advanced}.
\paragraph{Order of Permutation (OP)} 
The minimum number of compositions applied to get an identity mapping on the coordinate and component functions of a boolean mapping $S$ is called the order of permutation. OP is not an invariant property under affine equivalence \cite{nitaj2020new,abeer,cui2011improved,seghier2019advanced}.
\paragraph{Fixed Points (FP)} 
For a vectorial boolean mapping $S$, when an input value $(\varkappa)$ maps upon itself in the output, it is known as fixed point (FP). Mathematically:- 
\small
\begin{equation}
\label{eq:2}
S(\varkappa) = \varkappa
\end{equation}
\paragraph{Opposite Fixed Points (OFP)} 
In a vectorial boolean mapping $S$, if an input value $(\varkappa)$ maps on the inverse of itself in the output, it is known as opposite fixed point (OFP). Mathematically:-
\small
\begin{equation}
\label{eq:3}
S(\varkappa) = ({\varkappa})^\prime 
\end{equation}
If any $S$ has a high number of FPs or OFPs, it is considered to lack the desired randomness so they should kept to a minimum in order to avoid statistical attacks. This can be achieved with the application of an affine transformation, as FP and OFP are affine variant properties \cite{picek2016search}.
\paragraph{Bit Independence Criteria (BIC)}
A vectorial boolean mapping $S: F_2^n  \rightarrow F_2^m $ is said to satisfy BIC, if inverting an input bit $u$ has a changing effect on the output bit $v$ i.e. all the avalanche variables $(\gamma_u,\gamma_v,....)$ representing difference between bits of input and output of $S$ become pairwise independent \cite{webster1985design}. Mathematically, BIC is computed using correlation co-efficient of the avalanche variables:-
\small
\begin{equation}
\label{eq:4}
\rho(\gamma_u,\gamma_v)=\frac{cov(\gamma_u,\gamma_v)}{\sigma(\gamma_u)\sigma(\gamma_v)}
\end{equation}
Where; $\ u\neq v\ $ 
and $\ u,v = 1,2,.....,m\ $. 
The numerator in eq \ref{eq:4} denotes co-variance between two avalanche variables ($\gamma_u$ and $\gamma_v$), while $\sigma(\gamma_u)$ and $\sigma(\gamma_u)$ in the denominator is the product of the same variables. 
\renewcommand{\arraystretch}{1.5}
For the coordinate and component functions of $S$, the BIC value should stay in the range of $[0,1]$. An ideal S-box has a BIC value equal to $0$. If an affine transformation is applied to $S$, BIC does not stay invariant \cite{ao2017construction}.
\paragraph{Strict Avalanche Criteria (SAC)}
A vectorial boolean mapping $S$ is said to satisfy SAC such that whenever a single input bit $\varkappa$ is inverted, there is a $50$ percent probability of each of the output bit changing. SAC is satisfied if hamming weight $(wt)$ of all the coordinate and component functions of $S$ is $1$. Mathematically, it can be expressed in terms of directional derivative:-
\small
\begin{equation}
\label{eq:5}
     S_\gamma = S(\varkappa)\oplus S(\varkappa \oplus \gamma) 
\end{equation} 
Where $\gamma \in F_2^n, \gamma \neq 0, wt(\gamma)=1$. SAC is variant under affine transformation, which means that a function not satisfying SAC could be transformed to a function that satisfies SAC using affine equivalence \cite{saugdiccouglu2003cryptological}. 
\paragraph{Auto Correlation Transform (ACT)}
The ACT of $S$ is denoted by $\triangle_S (\gamma,\rho)$, where $\gamma \in F_2^n$, and $\rho \in F_2^m$. Then ACT of $S$ can be mathematically defined as follows:-
\small
\begin{equation}
\label{eq:11}
   \triangle_S (\gamma,\rho) = \sum_{\varkappa=0}^{2^{n-1}}(-1)^{\rho \cdot (S(\varkappa \oplus \gamma) \oplus S(\varkappa))} 
\end{equation}
ACT of $S$ is invariant under affine equivalence \cite{canteaut2021autocorrelations}.
\paragraph{Absolute Indicator (AI)}
The overall impact of auto correlation of a boolean function or a vectorial boolean function is studied by finding the maximum value in ACT spectrum, known as absolute indicator (AI). 
Let $f$ be a Boolean function over $F_2^n$ i.e. $f : F_2^n \rightarrow F_2$. The AI  of $f$ is defined as \cite{carlet2021boolean}:-
\small
\begin{equation}
\label{eq:12}
 \barwedge_f= \max_{\gamma \in F_2^n \backslash\{0\}} |\triangle_f (\gamma)|	
\end{equation}
Whereas, $\triangle_f (\gamma)$ is the autocorrelation of the function $f$ over $F_2^n$ shifted with $\gamma \in F_2^n$  and calculated as:-
\small
\begin{equation}
\label{eq:13}
F(D_\gamma f) = \triangle _f (\gamma) = \sum_{\varkappa \in F_2^n}(-1)^{D_\gamma f} 
\end{equation}
Here $D_\gamma f = f(\varkappa) \oplus f(\varkappa \oplus a)$.

For a vectorial boolean mapping $S: F_2^n \rightarrow F_2^m$, using (\ref{eq:11}) the AI of $S$ is defined as \cite{canteaut2019observations}:-
\small
\begin{equation}
\label{eq:14}
  \barwedge_S= \max_{\gamma, \rho \in F_2^n \backslash\{0\}} |\triangle_S (\gamma, \rho)|	
\end{equation}
$\barwedge_S$ stays invariant under affine transformation. For balanced vectorial boolean functions, the value of AI is always $0$. If a boolean mapping has high $\barwedge_S$, it becomes highly susceptible to cube attacks \cite{Dinur, carlet2021boolean}. 
\paragraph{Sum-of-Squares (SSI)}
The SSI of a vectorial boolean mapping $S$ is represented in terms of ACT as follows:-
\small
\begin{equation}
\label{eq:15}
   \doublebarwedge_S =\sum_{\gamma \in F_2^n, , \rho \in F_2^m} \triangle^2 _S (\gamma, \rho) 
\end{equation}  
SSI is not variant under affine equivalence \cite{li2023further}. For achieving good diffusion characteristics in a cryptographic algorithm, S-boxes should have low SSI and AI values \cite{zhang1996gac, carlet2021boolean}. 
\subsection{Linear Cryptanalysis}
The idea of this technique evolves around studying the non-linear layer of a block cipher to draw a linear approximations profile (LAT) between plaintext, key and the ciphertext to find correlations among them. Matsui introduced non-linearity as a significant criteria in the design of block ciphers \cite{matsui1993linear}. Block ciphers having a highly non-linear layer are resistant towards linear cryptanalysis. To study the number of active S-boxes in the linear trails of a block cipher, J. Daemon \textit{et al.} \cite{daemen2001wide} introduced linear branch number. 
Theoretically, in order to resist linear (and its variant) cryptanalysis techniques an S-box should have a low linear approximation probability, high non-linearity, high order of linear branch number, equal to none linear structures, and low order correlation immunity.
\paragraph{Linear Approximation Table (LAT)}
M. Matsui introduced linear cryptanalysis, as an effective way to define a cryptographic algorithm in terms of linear equations \cite{matsui1993linear, matsui1994first, heys2002tutorial}. A linear approximation table (LAT) is constructed by finding the probability of occurrence $P$ of all input ($\varrho$) and output ($\psi$) linear masks where $\varrho ,\psi \in F_2^n, F_2^m$ such that $\varrho \cdot \varkappa = \psi \cdot S(\varkappa)$. LAT can be mathematically represented as follows:-
\small
\begin{equation}
\label{eq:9}
\mathsf{LAT}(\varrho, \psi)= | \{\# \varkappa \ | \ \varrho \cdot \varkappa = \psi \cdot S(\varkappa)\} | -  2^{n-1}
\end{equation}
\begin{definition}
The maximum LAT table entry is measured in the form of Linear Approximation Probability (LAP), mathematically given as follows:-
\small
\begin{equation}
    \xi = \frac{1}{2^{n}}\max_{\varrho ,\psi \in F_2^n, F_2^m \backslash\{0\}} \bigg| \mathsf{LAT}(\varrho, \psi)\bigg|
\end{equation}

\end{definition}
The LAT table is of size $2^n \times 2^m$. The maximum LAT table entry and weight of coefficients in rows and columns stays is invariant under affine transformation, however the distribution of these entries is subject to change. Upper and lower bounds on all the entries of LAT are depicted in Table \ref{table:4}. 
\paragraph{Non-linearity (NL)}
Non-linearity is the measure of minimum hamming distance between the nonzero component functions of $S$ and their relative affine linear functions. NL of boolean a function $(f)$ is measured as follows:-
\small
\begin{equation}
\label{eq:6}
\mathsf{NL_f} = \min_{L(\varkappa) \in AL_n} dis(f(\varkappa), L(\varkappa))
\end{equation}

The NL of an $n \times m$ vectorial boolean mapping $S$ is calculated using walsh spectrum $\mathsf{W}(\gamma, \rho)$, where $\gamma, \rho \in F_2^n$:-

\small
\begin{equation}
\label {eq:7}
\mathsf{W}(\gamma,\rho)=\sum_{\varkappa\in F_2^n} (-1)^{(\gamma \cdot S(\varkappa) \oplus \rho\cdot \varkappa )}
\end{equation}

\small
\begin{equation}
\label {eq:8}
   \mathsf{NL_S} = 2^{n-1} - \frac{1}{2} \max_{\gamma \in F_2^n, \rho \in F_2^m} |\mathsf{W}(\gamma,\rho)|
\end{equation}
A high NL value is desirable in order to resist linear cryptanalysis. It is an affine invariant property of any mapping $S$ \cite{meier1989nonlinearity, waqas2014generation, carlet2010vectorial}.

\paragraph{Linear Branch Number (LBN)}
In block cipher design strategy, branch number is a significant property used to measure the strength of an S-box against linear cryptanalysis. An S-box with high branch number could increase the diffusion power of a cryptographic algorithm. Mathematically:-

\small
\begin{equation}
\label{eq:10}
    \mathsf{BN_L (S)}=\min_{\gamma \in F_2^n, \rho \in F_2^m\backslash\{0\}} \{w_t (\gamma) + w_t (\rho)\}   
\end{equation}

LBN is variant under affine transformation i.e. two S-boxes can be affine equivalent of each other but they can have different values of LBN \cite{sarkar2019relationship}. 

\paragraph{Linear Structures (LS)}
J.H. Evertse defines \cite{evertse1987linear}:
\begin{displayquote}
A block cipher has Linear Structures (LS) when there exist subsets of input, output and key bits, such that changing all the input and key bits has the same effect on the XOR of the output bits.
\end{displayquote}

\begin{definition}
Let $f: F_2^n \rightarrow F_2$ be a Boolean function, then for some $\gamma \in F_2^n,f(\varkappa \oplus \gamma) \oplus f(\varkappa) = c$ holds for all $\varkappa \in F_2^n$  where $c \in F_2$ is a constant, then $\gamma$ is called a LS of $f(\varkappa)$. More specifically, $\gamma$ is called an invariant LS of $f(\varkappa)$ if $c = 0$ and a complementary LS of $f(\varkappa)$ if $c = 1$ \cite{zhang1996gac}.
\end{definition}
\begin{definition}
Let $S: F_2^n \rightarrow F_2^m$ be a vectorial boolean mapping if for some $\gamma \in F_2^n, S(\varkappa \oplus \gamma) \oplus S(\varkappa) = c$  holds for all $\varkappa \in F_2^n$, where $c \in F_2^m$ is a constant, then $\gamma$ is called a LS of $S(\varkappa)$. More specifically, $\gamma$ is called an invariant LS of $S(\varkappa)$ if $c = 0$ and a complementary LS of $S(\varkappa)$ if $c = 1$ \cite{baksi2021default}.
\end{definition}

S-boxes having LS are vulnerable to attacks more robust than exhaustive key search, such as differential and linear cryptanalysis \cite{dobbertin2004survey}. LS is an affine variant property of any boolean mapping $S$ \cite{2020, carlet2021boolean}. 

\paragraph{Correlation Immunity (CI)}
CI is an independent statistical property of boolean functions depicting linear combinations between the input and output bits of $f$.

Similarly, a boolean mapping $S: F_2^n \rightarrow F_2^m$ is said to be $k^{th}$ order correlation immune (given $1 < k < n$), if all its coordinate functions $\gamma \cdot S(\varkappa)$ are independent of the linear combinations in its input. So using (\ref{eq:7}), we can say that $S$ is $k$-order CI when \cite{mukherjee2021design}:-

\small
\begin{equation}
\label{eq:16}
\{\mathsf{W}(\gamma, \rho) = 0: \ \gamma \in F_2^m\backslash\{0\}, \ \rho \in F_2^n,\ \forall \  1 \leq wt(\rho) \leq k \}
\end{equation}

CI shows variant behavior when an affine transformation is applied to $S$  \cite{carlet2021boolean}. For a good cryptographic S-box the lower is the order of CI the better. CI shares an inverse relation with NL, hence if CI holds a lower order the value of NL will be high, and vice versa.

\subsection{Differential Cryptanalysis}
Differential cryptanalysis is a technique that relies on exploiting chosen plaintext pairs and resulting ciphertext pairs, in order to gain the information of secret key. The pairs of plaintext and ciphertext are referred as differentials. And these differentials when combined for multiple rounds become differential characteristics of a cipher under study \cite{heys2002tutorial}. The probabilities of occurrence of certain output (ciphertext) differences corresponding fixed input (plaintext) differences are computed in the form of DDT. A low value of differential uniformity, high differential branch number, high value of propagation characteristics, and no undisturbed bits are a requirement for an S-box to resist against differential cryptanalysis and its variant attacks. 

\paragraph{Difference Distribution Table (DDT)}
Given two inputs: $\varkappa$ and $\varkappa'$ and their corresponding outputs: $\flat$ and $\flat'$, the respective input and output differences can be mathematically represented as:-

\small
\begin{equation}
\label{eq:17}
   \triangle_\varkappa = \varkappa \oplus \varkappa'	 
\end{equation}

\small
\begin{equation}
\label{eq:18}
  \triangle_\flat = \flat \oplus \flat'  
\end{equation}
Where $\varkappa, \varkappa' \in F_2^n$, $\flat, \flat' \in F_2^m$. Ideally, the probability that a particular output difference $(\triangle \flat)$ occurs given a particular input difference $(\triangle \varkappa)$ is $\frac{1}{2^{n-1}}$. Same as LAT, the lookup table of DDT is also of the dimension $2^n \times 2^m$. Mathematically:-

\small
\begin{equation}
\label{eq:19}
\mathsf{DDT(\triangle_\varkappa, \triangle_\flat)}= \# \{S(\varkappa \oplus  \triangle_\varkappa) \oplus S(\varkappa) = \triangle_\flat \}
\end{equation}

When an affine transformation is applied to $S$, DDT preserves the maximum entry and weight of coefficients in rows and columns, however their position is subject to a change  \cite{Boura_Canteaut_2018, bao2019peigen}. Table \ref{table:4} illustrates bounds on the possible values a single input difference can map to against a single output difference, except for $\triangle_\varkappa = 0$. Notably, the sum of all entries for a given $\triangle_\varkappa$ is upper bounded by $2^{n}$.

\begin{definition}
The maximum achievable value in the DDT of $S$ is called the differential uniformity (DU) of $S$. It can be mathematically represented as follows \cite{hasan2021c}:-

\small
\begin{equation}
\label{eq:20}
\mathsf{DU_S} = \max_{\triangle_\varkappa \neq 0, \triangle_\flat} \# \mathsf{DDT(\triangle_\varkappa, \triangle_\flat)}
\end{equation}
\end{definition}

For an S-box to resist differential cryptanalysis a smaller value of DU is required. If an affine transformation is applied to $S$. DU stays invariant \cite{bao2019peigen}.  

\paragraph{Differential Branch Number (DBN)}
The diffusion power of a vectorial boolean mapping $S$ is determined using differential branch number (DBN). It is measured by finding the minimum of hamming distance between two distinct inputs $\varkappa$, $\varkappa'$ and their corresponding outputs $S(\varkappa)$, $S(\varkappa')$. Mathematically \cite{sarkar2018bounds}:-

\small
\begin{equation}
\label{eq:21}
\mathsf{BN_D(S)} =\min_{\varkappa,\varkappa' \in F_2^n, \varkappa \neq \varkappa'} \{wt (\varkappa \oplus \varkappa' ) \oplus wt (S (\varkappa) \oplus S(\varkappa'))\} 
\end{equation}

An S-box having DBN greater than $2$ increases the chances of having more active S-boxes in iterative ciphers. DBN does not stay invariant under affine equivalence \cite{sarkar2019relationship}.  
 
\paragraph{Propagation Criteria (PC)}
A vectorial boolean function $S: F_2^n \rightarrow F_2^m$ is said to satisfy propagation criteria (PC) when $S_\gamma = S(\varkappa) \oplus S(\varkappa \oplus \gamma)$ is balanced given $\gamma \in F_2^n$. Mathematically $S$ satisfy $PC(l)$ iff \cite{carlet2021boolean, zhang1996gac, preneel1991propagation}:-

\small
\begin{equation}
\label{eq:23}
   \triangle_S (\gamma,\rho) = 0
\end{equation}

For $1 \leq wt(\gamma) \leq l$. 

$PC(l)$ of $S$ is not variant when an affine transformation is applied \cite{saugdiccouglu2003cryptological}. An S-box satisfies PC when $ACT_{max} = 0$. For an S-box to satisfy SAC, complying with PC is a pre-requisite \cite{carlet2021boolean}. S-boxes having low order $PC(l)$ have weak diffusion characteristics \cite{alvarez2012cryptographic}.

\paragraph{Undisturbed Bits (UDB)}
For a specific nonzero input difference $\triangle_\varkappa \in F_2^n$ of an S-box $S: F_2^n \rightarrow F_2^m$, some of the bits in the corresponding output difference $\triangle_\rho$ remain unchanged, these specific bits are referred to as undisturbed bits (UDB) \cite{makarim2014relating}.

Mathematically:-

\small
\begin{equation}
\label{eq:22}
\triangle_\flat = \{ \triangle_\rho = \rho_j \in F_2^m\ | Pr_S (\triangle_\varkappa \rightarrow \triangle_\rho) > 0 \}
\end{equation}

Here $j= (0,1,2,....m-1)$, if $\rho_j = d$, when $d \in F_2$ is fixed and $\triangle_\rho \in \triangle_\flat$ then $S$ has UDBs. UDB is invariant under affine transformation. An S-box should have no UDB in order to avoid the construction of truncated differentials. Currently, there is no theoretically proven upper bound on UDB. 

\subsection{Boomerang Cryptanalysis}
Boomerang attack is a technique of block cipher cryptanalysis based on the similar principle as differential cryptanlysis. However, instead of drawing one differential characteristic for the entire cipher, boomerang attack divides the cipher into two sub ciphers and generates two differentials characteristics also known as boomerang distinguishers. A low value of boomerang uniformity makes an S-box resistant towards boomerang attacks.

\paragraph{Boomerang Connectivity Table (BCT)}
The dependency between these two boomerang differentials can have an influence on their combined probability, hence it is of significance to observe the switching effect between the two. In order to achieve this purpose and simplify the analysis, C. Cid \textit{et al.}  proposed the Boomerang Connectivity Table (BCT) \cite{cid2018boomerang}. 

BCT of $S$ is given by a $2^n\times 2^m$ defined as $\beta T(\triangle_\varkappa,\nabla_\flat)$:-

\small
\begin{equation}
\begin{split}
\beta \mathsf{T}(\triangle_\varkappa,\nabla_\flat) = \# \{\varkappa \in F_2^{n} \big| \ S^{-1}  (S(\varkappa)\oplus \nabla_\flat )\oplus \\S^{-1} (S(\varkappa \oplus \triangle_\varkappa )\oplus \nabla_\flat)= \triangle_\varkappa\}
\end{split}
\end{equation}

$\triangle_\varkappa\in F_2^n =$ Input difference of S

$\nabla_\flat \in F_2^m =$ Output difference of S

\paragraph{Boomerang Uniformity (BU)}
The highest value in BCT is known as the boomerang uniformity (BU); where $[\triangle_\varkappa,\nabla_\flat] \neq 0$ \cite{boura2018boomerang}.
\small
\begin{equation}
\beta \mathsf{U} =\max_{\triangle_\varkappa,\nabla_\flat \in F_2^{n}, F_2^{m} \backslash\{0\}} \beta T(\triangle_\varkappa,\nabla_\flat )	
\end{equation}			
BCT and BU of an S-box are invariant under affine transformation \cite{carlet2021boolean, boura2018boomerang}.
\subsection{Differential-Linear Cryptanalysis}
Differential linear attack proposed by Hellman and Langford \cite{langford1994differential} is based on dividing a block cipher into two parts and then combining both differential and linear characteristics into one attack that covers the whole cipher. Hence, the complexity of a differential linear attack is primarily based on the dependency between the two parts.
\paragraph{Differential Linear Connectivity Table (DLCT)}
 Bar-On \textit{et al} \cite{bar2019dlct} proposed Differential Linear Connectivity Table (DLCT) in order to study the dependencies between two parts of the cipher so that better differential and linear attack characteristics could be drawn. The DLCT of $S$ is a $2^{n} \times 2^{m}$ table, where rows contain input differences and columns contain output bit masks.
For $\triangle_\varkappa \in F_2^{n}$ and $\rho \in F_2^{m}$, the DLCT entry $(\triangle_\varkappa,\rho)$ can be computed mathematically as follows:-
\small
\begin{equation}
\mathsf{DLCT}_S (\triangle_\varkappa,\rho) = \# \{\varkappa \in F_2^{n} \mid \ \rho \cdot S(\varkappa) = \rho \cdot S(\varkappa + \triangle_\varkappa)\} - 2^{n-1}		
\end{equation}
\paragraph{Differential Linear Uniformity (DLU)}
The differential linear uniformity (DLU) of $S$ can be defined by taking a maximum on $\mathsf{DLCT}_S (\triangle_\varkappa,\rho)$ for all nonzero values of $\triangle_\varkappa$ and $\rho$.  Mathematically:-
\small
\begin{equation}
\Gamma_S=\{\mathsf{DLCT}_S  (\triangle_\varkappa,\rho), \triangle_\varkappa \in F_2^{n} \backslash \{0\} \mbox{ and } \rho \in F_2^{m} \backslash \{0\}\}
\end{equation}
\begin{equation}
\mathsf{DLU}_S = \max (\Gamma_S)
\end{equation}
Under affine equivalence, DLCT and DLU are invariant \cite{JEONG2022101931, canteaut2019differential}. In order to resist against differential linear attack, an S-box should have low DLU.
\subsection{Algebraic Cryptanalysis}
The basic idea behind algebraic attacks involves converting the entire block cipher round function into a system of equations and then trying to solve these equations in order to deduce the secret key. Both, non-linear and linear components are transformed into polynomials. Unlike the linear component, the number of variables for nonlinear layer increases after every next round. Optimal S-boxes are better resistant towards algebraic crypltanalysis if they have a high value of algebraic degree and algebraic immunity, and high degree of interpolation polynomial. 
\paragraph{Algebraic Degree (AD)}
The polynomial representation of a boolean function $f: F_2^n \rightarrow F_2$, is known as its Algebraic Normal Form (ANF) \cite{carlet2021boolean}, written as:-
\small
\begin{equation}
\begin{split}
f(\varkappa) = c_0 \oplus c_1\varkappa_1 \oplus c_2\varkappa_2 \oplus . . . \oplus c_n\varkappa_n 
\oplus \\ c_{12} \varkappa_1 \varkappa_2 \oplus c_{13}\varkappa_1\varkappa_2 \oplus . . .\oplus c_{(n - 1)n} \varkappa_{(n - 1)} \varkappa_n\\ . . . \oplus c_{123.....n}\varkappa_1\varkappa_2\varkappa_3 . . . \varkappa_n
\end{split}
\end{equation}
Where the coefficients $c_i \in F_2$ form the elements of the truth table for ANF of $f(\varkappa)$. 
An n-bit vectorial boolean function $S: F_2^n \rightarrow F_2^m$ has $f_j = f_1, f_2, f_3,.....f_n$ coordinate functions and each can be uniquely represented in the form of ANF. 
\small
\begin{equation}
f_j (\varkappa) = \sum_{j = 1}^{{n}} C_j \prod_{j = 1}^{{n}} \varkappa_j
\end{equation}
AD is defined as the number of variables having non-zero coefficients in the highest product term of the ANF. Mathematically, for $S$, AD can be defined as:-
\small
\begin{equation}
    \mathsf{AD_S} = \max_{1 \leq j \leq n} {[deg(f_j(\varkappa))]}
\end{equation}

The AD is an affine invariant property of any mapping $S$ \cite{2020, carlet2021boolean}. S-boxess having a low AD are vulnerable to higher order differential attacks. AD has an inverse relation with CI, the higher is the order of CI, the lower is AD. 

\paragraph{Interpolation Polynomial (IP)}
In 1996, Jakobsen and Knudsen introduced interpolation attack on block ciphers \cite{jakobsen1997interpolation}. In this attack the round function or S-boxes are represented in the form of algebraic expressions. In either case, Lagrange interpolation technique is used to evaluate the algebraic expression. The Lagrange polynomial is constructed using as many plaintext/ciphertext pairs as the number of unknown coefficients in the algebraic representation. Attack complexity depends on number of terms or degree of the resulting polynomial expression. 

For any vectorial boolean mapping $S: F_2^n \rightarrow F_2^m$  with input $\varkappa_i= \{\varkappa_1,.....,\varkappa_n\}$ and output $\flat_i = \{\flat_1,.....,\flat_n\}$ where $i = 1, ......, n$, we find the Interpolation Polynomial (IP) $P_i(\varkappa)$ as follows:

\small
\begin{equation}
P_i (\varkappa) = \sum_{i=1}^{n} \flat_i \prod_{j=1; j\neq i}^{n} \frac{\varkappa - \varkappa_j}{\flat_i- \flat_j}
\end{equation}

IP is invariant under affine equivalence \cite{calvi2005lectures}, and a low degree of IP makes any S-box susceptible to interpolation attack.

\subsection{Side Channel Analysis}
Attacks on cryptographic implementations have evolved significantly since Kocher's first attempt to use physical leakages from a cryptosystem to extract sensitive information. Due to the increasing threat, designers are always trying to come up with efficient ways to resist against side channel attacks \cite{kocher2011introduction}. Starting from the design of components such as S-boxes with good cryptographic properties to resist side channels and moving up to the overall cipher level by employing countermeasures such as boolean or polynomial masking. Properties such as DPA SNR and transparency order are nowadays actively used to determine an S-box's resilience against DPA attacks. However, striking a perfect balance between the metrics related to traditional cryptanalysis and side channel analysis is an open research area. Currently, designers try to achieve cryptographic profiles that are well suited with the requirements of a particular infrastructure or standardization process (as in NIST LWC). In ideal terms, for an S-box to be resistant towards attacks such as DPA, it needs to posses a low DPA SNR and transparency order. 

\paragraph{DPA SNR}
Differential Power Analysis Signal to Noise Ratio (DPA SNR) was introduced by Sylvain G. \textit{et al}. in order to improve the quality of leakges from CMOS circuits by modelling ghost peaks \cite{guilley2004differential}. DPA SNR of $S: F_2^n \rightarrow F_2^m$ can be mathematically represented in form of walsh spectrum $\mathsf{W}(\gamma, \rho)$ (\ref{eq:7}), as follows:-  

\small
\begin{equation}
\mathsf{DPA_{SNR}(S)} = m2^{2n}\left ( \sum_{\gamma \in F_{2}^{n}} \left (\sum_{\rho \in F_{2}^{m}, wt(\rho)=1} \mathsf{W}(\gamma,\rho)  \right )^{4}\right )^{-\frac{1}{2}}
\end{equation}

As per the authors, DPA SNR increases when resistance of $S$ against linear and differential cryptanalysis increases and vice versa. Thus highlighting that alongside protected implementations, the choice of non-linear component also plays an important role in a block cipher's resistance against side channel attacks. DPA SNR of $S$ is variant under affine transformation \cite{picek2016search}.

\paragraph{Transparency Order (TO)}
Emmanuel P. \textit{et al}. extended the concept of (DPA SNR) to evaluate the leakages in a multi-bit DPA setup using hamming weight power model \cite{prouff2005dpa}. 

It is mathematically represented as follows:-
\small
\begin{equation}
\begin{split}
\mathsf{TO} = \max_{\beta \in F_{2}^{m}}(\big| m- 2 wt(\beta)\big| ) \times \frac{1}{2^{2n}-2^{n}} \times \\ 
\left ( \sum_{\gamma \in F_{2}^{n^{}}} \bigg| \sum_{\rho \in F_{2}^{m}, wt(u)=1} (-1) ^{\rho .\beta} \triangle_S (\gamma,\rho) \bigg|  \right )
\end{split}
\end{equation}

If an S-box has small value of TO, it is considered highly resistant against DPA attacks \cite{prouff2005dpa}. When affine transformation is applied to an S-box, TO does not stay invariant \cite{picek2016search}. 

\section{Discussion and Results}
\label{sec4}
We have given the details of our findings on Romulus, ASCON/ ISAP, Elephant, Photon-Beetle, and GIFT-COFB in Tables \ref{table:6}-\ref{table:7} against dedicated cryptographic profiles. We briefly explain them in subsequent subsections.
\subsection{Diffusion Metrics}
If we look at the basic avalanche characteristics of all these S-boxes they satisfy the criteria of balancedness. This helps reduce the chances of getting any high probability linear approximations, in turn assuring the resistance of these S-boxes towards linear cryptanalysis \cite{farwa2016highly}. The S-boxes under analysis are bijective, so they satisfy the criteria of permutation. From the bounds discussed in Table \ref{table:4}, we have developed an understanding that ideally the S-boxes should have a full cycle length for order of permutation. Presumably, S-boxes having short iterative periods are vulnerable to attacks. In order to avoid such attacks designers can use affine equivalence to improve the order of permutation of these S-boxes \cite{nitaj2020new, abeer, cui2011improved, seghier2019advanced}.\\
Out of the six, Ascon/ ISAP S-box has a FP at $\varkappa= 255$. Photon-Beetle/ Elephant/ GIFT-COFB, each has one OFPs at $\varkappa= 14/ 8/ 3$. S-boxes having FPs can alter the overall diffusion power that an iterative cipher has to offer and make it vulnerable to invariant subspace attacks, it may be rectified using affine equivalent S-box. \\
The upper bounded BIC value for all the S-boxes indicates that the input and output avalanche vectors are fully correlated. \\
Subsequently, none of the S-box satisfies the principles of completeness and avalanche depicted by their SAC failure. The AI values for all the S-boxes achieve the upper bound, and the values for that of SSI seem a bit better but do not satisfy the lower bound requirement.  As discussed earlier, for an S-box to satisfy SAC, one of the prerequisites is compliance with PC. In Tables \ref{table:6}-\ref{table:7} we can see that all S-boxes have PC $= 0$, which again points towards the weak avalanche profile. These findings indicate that an attacker can utilize this information to mount statistical attacks on the underlying cipher \cite{webster1985design, singh2017analysis}.  
\subsection{Linear Metrics}
Security towards linear cryptanalysis attacks is ensured when an S-box has a low LAP and higher non linearity. Our results show that all $6$ S-boxes under analysis have achieved the desired bounds for both properties. The LBN of all $6$ S-boxes is closer to the LB, which is opposite to the ideal requirements. However, if the design requires it, LBN can be improved using affine transformations. The presence of LS makes it possible for the cipher to leak significant amount of information if specific group(s) of bits in the input are flipped. All the S-boxes are first order correlation immune, which makes them secure against zero-correlation attacks. 
\subsection{Differential and Boomerang Metrics}
Almost Perfect Nonlinear (APN) S-boxes, in odd dimensions are differentially $2$ uniform \cite{abeer}. The only odd dimension case in S-boxes under study is of Ascon/ ISAP, and since its has a DU: $8$. The remaining S-boxes have a DU within the desired bounds.  The DBN of Elephant and Photon-Beetle S-box satisfies the UB,  Ascon/ ISAP is closer to the UB, and for Romulus and GIFT-COFB DBN value stands on the LB. A good DBN ensures more number of active S-boxes across multiple rounds in block ciphers. The UDB are detected in all $6$ S-boxes. In \cite{makarim2014relating} authors have extensively described how the presence of UDB in an S-box impacts other cryptographic properties, such as the occurrence of LS, as we can see in Tables \ref{table:6}-\ref{table:7}.\\
Tezcan et al. used these UDB and LS to devise 4 and 5 round impossible, truncated and improbab differentials, and break 5 out of 12 Ascon rounds \cite{tezcan2016truncated}. \\
In regards to boomerang cryptanalysis, the BU for all S-boxes is attaining the UB, which is opposite to the required bounds in order to be secure against Boomerang attacks.

\subsection{Differential-Linear Metrics}
Combined attacks make it easier for an attacker to break a cipher by defining short differential and linear characteristics, as in the case of differential-linear cryptanalysis. Hence, attaining the LB on DLU is crucial to gain resistance against these attacks. In case of S-boxes under analysis, we can observe from Tables \ref{table:6}-\ref{table:7}, that the DLU is a little way up from the LB, hence paving a way for differential-linear attack with a reduced complexity. 

\subsection{Algebraic Metrics}
Previously, we have talked about the inverse relation between CI and AD, and we observed that all $6$ S-boxes have a CI $=0$. So given the inverse relation, AD of all S-boxes should be obtaining the UB, which in our case is true. AD determines the complexity of an S-box's ANF, the higher the degree is the more complex it is to launch an algebraic attack on a cipher. S-boxes like those of Ascon/ ISAP having an AD $=2$ are referred to as quadratic and those having AD $=3$ are known as cubic. The functions that posses and AD $=1$ are known as affine.  
\subsection{Side Channel Metrics}
During design and analysis phase, DPA SNR and TO of S-box are the two metrics that can be used to determine a cipher's resistance towards SCA (DPA) attacks. From the conventional cryptanalysis perspective, S-boxes having a high NL and low DU are prone to SCA attacks and have a high resistance towards differential and linear cryptanalysis, and vice versa. Hence, finding the ideal balance between the two axes of attack metrics is crucial to the security of a lightweight block cipher. 
Since the S-boxes of LWC finalists vary in size (4, 5 and 8 bits), a direct comparison of the results we obtained of their side-channel metrics is not fair, since the required optimal bounds for each size are different (see Table \ref{table:4}). Hence to draw a meaningful comparison, we normalize the DPA SNR and TO using the ideal or worse case bounds as a reference point.
\begin{itemize}
    \item DPA SNR can be normalized using the identity-case bound \cite{guilley2004differential}, as follows:
    \small
    \begin{equation}
    \mathsf{DPA_{SNR}(I) = \sqrt{m}}
    \end{equation}
    Here, in the case of a balanced S-box $m$ is the number of input/output bits. We now define the normalized DPA SNR as follows:-
     \small
    \begin{equation}
    \mathsf{DPA_{SNR}(norm) = \frac{{DPA_{SNR}(S)} - {DPA_{SNR}(I)} }{DPA_{SNR}(S)}}
    \end{equation}
    
   The results of the suggested normalization are in the range $[0,1]$. Here, a value $0$ indicates that the S-box under analysis has a DPA SNR equal to that of an identity S-box, which theoretically is the best DPA SNR possible for an S-box to be resistant towards DPA attacks. For example, in the case of 8-bit S-box, the identity has a DPA SNR $2.82$ and the DPA SNR of Romulus is $6.312455$. The normalized DPA SNR in the 8-bit case is $0.552$, suggesting that Romulus S-box potentially leaks $55.2\%$ more than the identity S-box. In the case of 4-bit sbox, the identity has a DPA SNR $2$, while GIFT, PHOTON and Elephant have DPA SNR $2.128608$ and $2.398501$, respectively. The normalized DPA SNR in the case of GIFT is $0.064304$, and in the case of PHOTON and Elephant is $0.1992505$, respectively. The normalization in the case of 4-bit S-boxes suggests GIFT has the lowest leakage amongst the three 4-bit S-boxes, as it leaks only $6.4304\%$ more than the 4-bit identity S-box.
    \item TO can be normalized using the worse-case (highest leakage) bound \cite{prouff2005dpa}, as follows:-
    \begin{equation}
    \mathsf{{TO}_{highest-leakage} = m}
    \end{equation}
    Here, in the case of a balanced S-box $m$ is the number of input/output bits. Based on the worse-case bound, we now define the normalized TO as follows:-
    \small
    \begin{equation}
    \mathsf{TO(norm) = \frac{TO_{highest-leakage} - {TO_{S}} } {TO_{highest-leakage}}}
    \end{equation}
    The results of this normalization lie in the range [0,1], where a factor 0 depicts that the TO is the same as the worst-case scenario, i.e., the maximum TO value, implying the highest leakage in the case of DPA attacks. Any value greater than $0$ suggests relative improvement in the TO value and hence reduced leakage. For example, in the case of 8-bit S-box, the worse-case TO is $8$ and the TO of Romulus is $7.174510$. The normalized TO in the 8-bit case is $0.1031$, suggesting that Romulus S-box potentially leaks $10.31\%$ less than the worse-case S-box.
\end{itemize}




\section{Exploring Trade-offs in S-box Design}
\label{sec5}
Based on our analysis in Section \ref{sec4}, several finalist S-boxes exhibit weaknesses in classical cryptographic metrics — such as fixed points, differential uniformity, non-linearity, linear structures, etc. So to explore the possibility of improving the cryptographic robustness of S-boxes beyond those used in the NIST LWC finalists, we generated a new $4\times4$ bit test S-box using the PEIGEN  \cite{peigen} tool for S-box evaluation and generation, as shown in Table \ref{table:8}, using the following constraints:-
\begin{itemize}
    \item Permutation: \checkmark
    \item Differential Uniformity: 4
    \item Algebraic Degree: 3
    \item Linear Structures: 0
    \item Implementation Cost: 8
\end{itemize}

In PEIGEN, cost is a composite metric that captures the implementation efficiency in terms of both hardware and software, focusing on the following traits:-
\begin{itemize}
\item Gate Equivalent (GE): refers to a 2-input NAND gate unit,
\item Bitslice Gate: for software implementations (optimized 32-bits),
\item Circuit depth in terms of latency (depth vs area), and
\item Multiplicative complexity: minimizing number of non-linear operations (for side channel secure ciphers where masking and other countermeasures are involved). 

\end{itemize}

\renewcommand{\arraystretch}{1.5}
\begin{table*}[h]
\centering
\caption{Proposed 4-bit S-box}
\label{table:8}
\begin{tabular}{l|llllllllllllllll} 
\hline
4-bit & 0 & 4 & 5 & 1 & c & 9 & 6 & 7 & 3 & a & e & 8 & b & 2 & f & d \\ 
\hline
\end{tabular}
\end{table*}

Fixing the implementation cost ensures that the resultant S-box implementation lies within the acceptable practical bounds for lightweight cryptography implementations, while retaining the desired cryptographic profile.

\subsection{Comparison and Analysis}
Table \ref{table:7}, highlights the comparison between the S-box new 4-bit S-box and  and the 4-bit S-boxes from GIFT-COFB, Photon-Beetle and Elephant. From the classical cryptanalysis perspective all four S-boxes have the same differential uniformity, non-linearity, linear approximation, algebraic degree and correlation immunity. A noticeable difference occurs in the properties of the new S-box when compared to the other three, as follows:-

    - The values for BIC and SAC are better.
    
    - No linear structures and undisturbed bits.
    
    - Boomerang uniformity has improved.
    
\noindent
Some of the trade-offs that exist in the properties of the new S-box are as follows:-
    
    - $3$ fixed points are present in the new S-box while the others have null.
    
    - DPA-SNR value of $2.5786$ is a little higher than for all three finalists, indicating slightly weaker resistance to side-channel (DPA) leakage, but offers optimal resistance against conventional cryptanalysis.
    
    - TO of $3.4$ is a little higher than that of Elephant ($3.26$), but a little lower than that of PHOTON and GIFT-COFB, indicating a reasonable balance between leakage and classical security.

The current findings underscore that even when the classical cryptanalysis properties are optimized within practical bounds, striking a perfect balance with the side-channel metrics may not be possible. In practice, identifying such balanced S-boxes remains a trial-and-error process that involves testing multiple generation strategies and parameter choices until an acceptable trade-off is found.
Moreover, it is important to emphasize that the S-box alone does not define the overall security of a lightweight block cipher. Resistance to differential and linear cryptanalysis, as well as to side-channel analysis, also depends heavily on other design aspects such as the diffusion layer, round constants, and key schedule, etc. A strong S-box is therefore only one building block of a well-rounded cryptographic primitive.

\section{Scope and Limitations}
\label{sec6}

Our research presents a comprehensive evaluation of the cryptographic properties of S-boxes used in finalist ciphers of the NIST LWC competition. For the sake of clarity and accuracy it is important that we acknowledge the scope and limitations of our analysis. 

Firstly, we would like to clarify that our analysis of S-boxes is conducted from a theoretical and metric-based perspective. We do not claim to perform any practical side-channel attacks (e.g., DPA, SPA) or cryptanalytic attacks (e.g., differential, linear) on the full/reduced-rounds structure of NIST LWC finalists.

Strong S-boxes contribute significantly to overall cipher security. We have evaluated the S-boxes individually, based on well-established cryptographic metrics listed in Table \ref{table:3}. Such compact analysis is a crucial first-step in determining the strength of the confusion layer when designing a new cipher, but it should be noted that full cipher-level cryptanalysis and side-channel analysis requires holistic evaluation -- involving linear layers, modes of encryption, masking strategies, implementation environments, and leakage models. These are beyond the scope of this study and can be seen as potential future work.

Therefore, we specifically refrain from extrapolating our S-box component-level findings to make any assertions on the overall security strength of the cipher.

\section{Conclusion}
\label{sec7}
Since the announcement of NIST lightweight cryptography competition, a considerable number of attacks have been published for exploiting the vulnerabilities in the non-linear component of the competitors under consideration. This paper presents a detailed cryptographic security analysis of the S-boxes of six finalists, examining the $24$ cryptographic properties that are relevant to established attack methods (see\ref{sec3}). These six S-boxes exhibit characteristics such as high SAC and BIC values, degree-2 polynomials, and undisturbed bits — that may increase the susceptibility to known differential, linear, and boomerang attacks. The existence of degree 2 polynomials is the primary source of linear structures, followed by undisturbed bits. Improving the S-box design may, therefore, help improve their overall security margin of these offered by these permutations.

\section{Declarations}
\label{sec8}

\textbf{Author Contribution Statement}:

All authors contributed equally to the writing of the manuscript. 

\textbf{Author Competing Interest statement}:
 Sajid Ali Khan has position on editorial board of Discover Computing, and was not involved in the review or decisions related to this manuscript. The other authors declare that they have no known competing financial interests or personal relationships that could have appeared to influence the work reported in this paper.

\textbf{Funding Declaration}:

Not applicable

\textbf{Ethics and Consent to Participate declarations}:

Not applicable

\textbf{Consent for publication Declaration}:

Not applicable

\textbf{Ethics, Consent to Participate, and Consent to Publish declarations}:
Not Applicable

\textbf{Availability of Data}:
The data that support the findings of this study are available from the corresponding author, upon request.

\textbf{Preprint}:
A preprint of this article is available at \cite{naseer2024s}


\begin{thebibliography}{131}
\ifx \bisbn   \undefined \def \bisbn  #1{ISBN #1}\fi
\ifx \binits  \undefined \def \binits#1{#1}\fi
\ifx \bauthor  \undefined \def \bauthor#1{#1}\fi
\ifx \batitle  \undefined \def \batitle#1{#1}\fi
\ifx \bjtitle  \undefined \def \bjtitle#1{#1}\fi
\ifx \bvolume  \undefined \def \bvolume#1{\textbf{#1}}\fi
\ifx \byear  \undefined \def \byear#1{#1}\fi
\ifx \bissue  \undefined \def \bissue#1{#1}\fi
\ifx \bfpage  \undefined \def \bfpage#1{#1}\fi
\ifx \blpage  \undefined \def \blpage #1{#1}\fi
\ifx \burl  \undefined \def \burl#1{\textsf{#1}}\fi
\ifx \doiurl  \undefined \def \doiurl#1{\url{https://doi.org/#1}}\fi
\ifx \betal  \undefined \def \betal{\textit{et al.}}\fi
\ifx \binstitute  \undefined \def \binstitute#1{#1}\fi
\ifx \binstitutionaled  \undefined \def \binstitutionaled#1{#1}\fi
\ifx \bctitle  \undefined \def \bctitle#1{#1}\fi
\ifx \beditor  \undefined \def \beditor#1{#1}\fi
\ifx \bpublisher  \undefined \def \bpublisher#1{#1}\fi
\ifx \bbtitle  \undefined \def \bbtitle#1{#1}\fi
\ifx \bedition  \undefined \def \bedition#1{#1}\fi
\ifx \bseriesno  \undefined \def \bseriesno#1{#1}\fi
\ifx \blocation  \undefined \def \blocation#1{#1}\fi
\ifx \bsertitle  \undefined \def \bsertitle#1{#1}\fi
\ifx \bsnm \undefined \def \bsnm#1{#1}\fi
\ifx \bsuffix \undefined \def \bsuffix#1{#1}\fi
\ifx \bparticle \undefined \def \bparticle#1{#1}\fi
\ifx \barticle \undefined \def \barticle#1{#1}\fi
\bibcommenthead
\ifx \bconfdate \undefined \def \bconfdate #1{#1}\fi
\ifx \botherref \undefined \def \botherref #1{#1}\fi
\ifx \url \undefined \def \url#1{\textsf{#1}}\fi
\ifx \bchapter \undefined \def \bchapter#1{#1}\fi
\ifx \bbook \undefined \def \bbook#1{#1}\fi
\ifx \bcomment \undefined \def \bcomment#1{#1}\fi
\ifx \oauthor \undefined \def \oauthor#1{#1}\fi
\ifx \citeauthoryear \undefined \def \citeauthoryear#1{#1}\fi
\ifx \endbibitem  \undefined \def \endbibitem {}\fi
\ifx \bconflocation  \undefined \def \bconflocation#1{#1}\fi
\ifx \arxivurl  \undefined \def \arxivurl#1{\textsf{#1}}\fi
\csname PreBibitemsHook\endcsname

\bibitem[\protect\citeauthoryear{Subramanian et~al.}{2017}]{RezaMehran}
\begin{barticle}
\bauthor{\bsnm{Subramanian}, \binits{S.}},
\bauthor{\bsnm{Mozaffari-Kermani}, \binits{M.}},
\bauthor{\bsnm{Azarderakhsh}, \binits{R.}},
\bauthor{\bsnm{Nojoumian}, \binits{M.}}:
\batitle{Reliable hardware architectures for cryptographic block ciphers led and hight}.
\bjtitle{IEEE Transactions on Computer-Aided Design of Integrated Circuits and Systems}
\bvolume{36}(\bissue{10}),
\bfpage{1750}--\blpage{1758}
(\byear{2017})
\doiurl{10.1109/TCAD.2017.2661811}
\end{barticle}
\endbibitem

\bibitem[\protect\citeauthoryear{Bassham et~al.}{2018}]{bassham2018submission}
\begin{botherref}
\oauthor{\bsnm{Bassham}, \binits{L.}},
\oauthor{\bsnm{{\c{C}}al{\i}k}, \binits{{\c{C}}.}},
\oauthor{\bsnm{McKay}, \binits{K.}},
\oauthor{\bsnm{Turan}, \binits{M.S.}}:
Submission requirements and evaluation criteria for the lightweight cryptography standardization process.
US National Institute of Standards and Technology
(2018)
\end{botherref}
\endbibitem

\bibitem[\protect\citeauthoryear{Turan et~al.}{2021}]{turan2021status}
\begin{botherref}
\oauthor{\bsnm{Turan}, \binits{M.S.}},
\oauthor{\bsnm{McKay}, \binits{K.}},
\oauthor{\bsnm{Chang}, \binits{D.}},
\oauthor{\bsnm{Calik}, \binits{C.}},
\oauthor{\bsnm{Bassham}, \binits{L.}},
\oauthor{\bsnm{Kang}, \binits{J.}},
\oauthor{\bsnm{Kelsey}, \binits{J.}}, et al.:
Status report on the second round of the nist lightweight cryptography standardization process.
National Institute of Standards and Technology Internal Report
\textbf{8369}
(2021)
\end{botherref}
\endbibitem

\bibitem[\protect\citeauthoryear{Division}{}]{finalist}
\begin{botherref}
\oauthor{\bsnm{Division}, \binits{C.S.}}:
Finalists - lightweight cryptography: Csrc.
CSRC
\end{botherref}
\endbibitem

\bibitem[\protect\citeauthoryear{Turan et~al.}{2023}]{turan2023status}
\begin{botherref}
\oauthor{\bsnm{Turan}, \binits{M.S.}},
\oauthor{\bsnm{McKay}, \binits{K.}},
\oauthor{\bsnm{Chang}, \binits{D.}},
\oauthor{\bsnm{Kang}, \binits{J.}},
\oauthor{\bsnm{Waller}, \binits{N.}},
\oauthor{\bsnm{Kelsey}, \binits{J.M.}},
\oauthor{\bsnm{Bassham}, \binits{L.E.}},
\oauthor{\bsnm{Hong}, \binits{D.}}:
Status report on the final round of the nist lightweight cryptography standardization process.
NIST
(2023)
\end{botherref}
\endbibitem

\bibitem[\protect\citeauthoryear{CSRC}{2023}]{result}
\begin{botherref}
\oauthor{\bsnm{CSRC}, \binits{N.-}}:
Announcing lightweight cryptography selection | $csrc_2023$.
NIST | CSRC
(2023)
\end{botherref}
\endbibitem

\bibitem[\protect\citeauthoryear{Tezcan}{2020}]{cihangir1}
\begin{botherref}
\oauthor{\bsnm{Tezcan}, \binits{C.}}:
Analysis of ascon, drygascon, and shamash permutations.
Cryptology ePrint Archive, Report 2020/1458
(2020)
\end{botherref}
\endbibitem

\bibitem[\protect\citeauthoryear{Li et~al.}{2017}]{li2017cryptanalysis}
\begin{barticle}
\bauthor{\bsnm{Li}, \binits{Y.}},
\bauthor{\bsnm{Zhang}, \binits{G.}},
\bauthor{\bsnm{Wang}, \binits{W.}},
\bauthor{\bsnm{Wang}, \binits{M.}}:
\batitle{Cryptanalysis of round-reduced ascon}.
\bjtitle{Science China Information Sciences}
\bvolume{60}(\bissue{3}),
\bfpage{1}--\blpage{2}
(\byear{2017})
\end{barticle}
\endbibitem

\bibitem[\protect\citeauthoryear{Dobraunig et~al.}{2015}]{dobaruyng}
\begin{botherref}
\oauthor{\bsnm{Dobraunig}, \binits{C.}},
\oauthor{\bsnm{Eichlseder}, \binits{M.}},
\oauthor{\bsnm{Mendel}, \binits{F.}},
\oauthor{\bsnm{Schl{\"a}ffer}, \binits{M.}}:
Cryptanalysis of ascon.
Topics in Cryptology --- CT-RSA 2015,
371--387
(2015)
\end{botherref}
\endbibitem

\bibitem[\protect\citeauthoryear{Duarte-Sanchez and Halak}{2021}]{Duarte-Sanchez2021}
\begin{bbook}
\bauthor{\bsnm{Duarte-Sanchez}, \binits{J.E.}},
\bauthor{\bsnm{Halak}, \binits{B.}}:
In: \beditor{\bsnm{Halak}, \binits{B.}} (ed.)
\bbtitle{A Cube Attack on a Trojan-Compromised Hardware Implementation of Ascon},
pp. \bfpage{69}--\blpage{88}.
\bpublisher{Springer},
\blocation{Cham}
(\byear{2021}).
\doiurl{10.1007/978-3-030-62707-2_2}
\end{bbook}
\endbibitem

\bibitem[\protect\citeauthoryear{Erlacher et~al.}{2022}]{Erlacher}
\begin{barticle}
\bauthor{\bsnm{Erlacher}, \binits{J.}},
\bauthor{\bsnm{Mendel}, \binits{F.}},
\bauthor{\bsnm{Eichlseder}, \binits{M.}}:
\batitle{Bounds for the security of ascon against differential and linear cryptanalysis}.
\bjtitle{IACR Transactions on Symmetric Cryptology}
\bvolume{2022}(\bissue{1}),
\bfpage{64}--\blpage{87}
(\byear{2022})
\doiurl{10.46586/tosc.v2022.i1.64-87}
\end{barticle}
\endbibitem

\bibitem[\protect\citeauthoryear{El~Hirch et~al.}{2022}]{ElHirch_Mella}
\begin{barticle}
\bauthor{\bsnm{El~Hirch}, \binits{S.}},
\bauthor{\bsnm{Mella}, \binits{S.}},
\bauthor{\bsnm{Mehrdad}, \binits{A.}},
\bauthor{\bsnm{Daemen}, \binits{J.}}:
\batitle{Improved differential and linear trail bounds for ascon}.
\bjtitle{IACR Transactions on Symmetric Cryptology}
\bvolume{2022}(\bissue{4}),
\bfpage{145}--\blpage{178}
(\byear{2022})
\doiurl{10.46586/tosc.v2022.i4.145-178}
\end{barticle}
\endbibitem

\bibitem[\protect\citeauthoryear{Leander et~al.}{2018}]{Leander}
\begin{barticle}
\bauthor{\bsnm{Leander}, \binits{G.}},
\bauthor{\bsnm{Tezcan}, \binits{C.}},
\bauthor{\bsnm{Wiemer}, \binits{F.}}:
\batitle{Searching for subspace trails and truncated differentials}.
\bjtitle{IACR Transactions on Symmetric Cryptology}
\bvolume{2018}(\bissue{1}),
\bfpage{74}--\blpage{100}
(\byear{2018})
\doiurl{10.13154/tosc.v2018.i1.74-100}
\end{barticle}
\endbibitem

\bibitem[\protect\citeauthoryear{Baksi et~al.}{2021}]{9474092}
\begin{botherref}
\oauthor{\bsnm{Baksi}, \binits{A.}},
\oauthor{\bsnm{Breier}, \binits{J.}},
\oauthor{\bsnm{Chen}, \binits{Y.}},
\oauthor{\bsnm{Dong}, \binits{X.}}:
Machine learning assisted differential distinguishers for lightweight ciphers.
2021 Design, Automation \& Test in Europe Conference \& Exhibition (DATE),
176--181
(2021)
\doiurl{10.23919/DATE51398.2021.9474092}
\end{botherref}
\endbibitem

\bibitem[\protect\citeauthoryear{Dobraunig et~al.}{2015}]{Christoph2015}
\begin{botherref}
\oauthor{\bsnm{Dobraunig}, \binits{C.}},
\oauthor{\bsnm{Eichlseder}, \binits{M.}},
\oauthor{\bsnm{Mendel}, \binits{F.}}:
Heuristic tool for linear cryptanalysis with applications to caesar candidates.
Advances in Cryptology -- ASIACRYPT 2015,
490--509
(2015)
\end{botherref}
\endbibitem

\bibitem[\protect\citeauthoryear{Kannwischer et~al.}{2020}]{Kannwischer}
\begin{barticle}
\bauthor{\bsnm{Kannwischer}, \binits{M.J.}},
\bauthor{\bsnm{Pessl}, \binits{P.}},
\bauthor{\bsnm{Primas}, \binits{R.}}:
\batitle{Single-trace attacks on keccak}.
\bjtitle{IACR Transactions on Cryptographic Hardware and Embedded Systems}
\bvolume{2020}(\bissue{3}),
\bfpage{243}--\blpage{268}
(\byear{2020})
\doiurl{10.13154/tches.v2020.i3.243-268}
\end{barticle}
\endbibitem

\bibitem[\protect\citeauthoryear{Udvarhelyi et~al.}{2021}]{udvarhelyi2021security}
\begin{botherref}
\oauthor{\bsnm{Udvarhelyi}, \binits{B.}},
\oauthor{\bsnm{Bronchain}, \binits{O.}},
\oauthor{\bsnm{Standaert}, \binits{F.-X.}}:
Security analysis of deterministic re-keying with masking and shuffling: Application to isap.
International Workshop on Constructive Side-Channel Analysis and Secure Design,
168--183
(2021).
Springer
\end{botherref}
\endbibitem

\bibitem[\protect\citeauthoryear{CUI et~al.}{2022}]{yaxin}
\begin{barticle}
\bauthor{\bsnm{CUI}, \binits{Y.}},
\bauthor{\bsnm{XU}, \binits{H.}},
\bauthor{\bsnm{QI}, \binits{W.}}:
\batitle{Milp-based linear attacks on round-reduced gift}.
\bjtitle{Chinese Journal of Electronics}
\bvolume{31}(\bissue{1}),
\bfpage{89}--\blpage{98}
(\byear{2022})
{\href{https://arxiv.org/abs/https://ietresearch.onlinelibrary.wiley.com/doi/pdf/10.1049/cje.2020.00.113}{{https://ietresearch.onlinelibrary.wiley.com/doi/pdf/10.1049/cje.2020.00.113}}}
\end{barticle}
\endbibitem

\bibitem[\protect\citeauthoryear{Cao and Zhang}{2019}]{Meichun}
\begin{barticle}
\bauthor{\bsnm{Cao}, \binits{M.}},
\bauthor{\bsnm{Zhang}, \binits{W.}}:
\batitle{Related-key differential cryptanalysis of the reduced-round block cipher gift}.
\bjtitle{IEEE Access}
\bvolume{7},
\bfpage{175769}--\blpage{175778}
(\byear{2019})
\doiurl{10.1109/ACCESS.2019.2957581}
\end{barticle}
\endbibitem

\bibitem[\protect\citeauthoryear{Eskandari et~al.}{2019}]{Eskandri}
\begin{botherref}
\oauthor{\bsnm{Eskandari}, \binits{Z.}},
\oauthor{\bsnm{Kidmose}, \binits{A.B.}},
\oauthor{\bsnm{K{\"o}lbl}, \binits{S.}},
\oauthor{\bsnm{Tiessen}, \binits{T.}}:
Finding integral distinguishers with ease.
Selected Areas in Cryptography -- SAC 2018,
115--138
(2019)
\end{botherref}
\endbibitem

\bibitem[\protect\citeauthoryear{Ji et~al.}{2020}]{ji2020improved}
\begin{botherref}
\oauthor{\bsnm{Ji}, \binits{F.}},
\oauthor{\bsnm{Zhang}, \binits{W.}},
\oauthor{\bsnm{Zhou}, \binits{C.}},
\oauthor{\bsnm{Ding}, \binits{T.}}:
Improved (related-key) differential cryptanalysis on gift.
International Conference on Selected Areas in Cryptography,
198--228
(2020).
Springer
\end{botherref}
\endbibitem

\bibitem[\protect\citeauthoryear{Sun et~al.}{2021}]{Sun_Wang_Wang_2021}
\begin{barticle}
\bauthor{\bsnm{Sun}, \binits{L.}},
\bauthor{\bsnm{Wang}, \binits{W.}},
\bauthor{\bsnm{Wang}, \binits{M.}}:
\batitle{Linear cryptanalyses of three aeads with gift-128 as underlying primitives}.
\bjtitle{IACR Transactions on Symmetric Cryptology}
\bvolume{2021}(\bissue{2}),
\bfpage{199}--\blpage{221}
(\byear{2021})
\end{barticle}
\endbibitem

\bibitem[\protect\citeauthoryear{Zhu et~al.}{2018}]{cryptoeprint:2018/390}
\begin{botherref}
\oauthor{\bsnm{Zhu}, \binits{B.}},
\oauthor{\bsnm{Dong}, \binits{X.}},
\oauthor{\bsnm{Yu}, \binits{H.}}:
Milp-based differential attack on round-reduced gift.
Cryptology ePrint Archive, Paper 2018/390
(2018).
\url{https://eprint.iacr.org/2018/390}
\end{botherref}
\endbibitem

\bibitem[\protect\citeauthoryear{Inoue and Minematsu}{2021}]{cryptoeprint:2021/737}
\begin{botherref}
\oauthor{\bsnm{Inoue}, \binits{A.}},
\oauthor{\bsnm{Minematsu}, \binits{K.}}:
Gift-cofb is tightly birthday secure with encryption queries.
Cryptology ePrint Archive, Paper 2021/737
(2021).
\url{https://eprint.iacr.org/2021/737}
\end{botherref}
\endbibitem

\bibitem[\protect\citeauthoryear{Jang et~al.}{2020}]{cryptoeprint:2020/1405}
\begin{botherref}
\oauthor{\bsnm{Jang}, \binits{K.}},
\oauthor{\bsnm{Kim}, \binits{H.}},
\oauthor{\bsnm{Eum}, \binits{S.}},
\oauthor{\bsnm{Seo}, \binits{H.}}:
Grover on gift.
Cryptology ePrint Archive, Paper 2020/1405
(2020).
\url{https://eprint.iacr.org/2020/1405}
\end{botherref}
\endbibitem

\bibitem[\protect\citeauthoryear{Picek et~al.}{2014}]{6855573}
\begin{botherref}
\oauthor{\bsnm{Picek}, \binits{S.}},
\oauthor{\bsnm{Ege}, \binits{B.}},
\oauthor{\bsnm{Papagiannopoulos}, \binits{K.}},
\oauthor{\bsnm{Batina}, \binits{L.}},
\oauthor{\bsnm{Jakobović}, \binits{D.}}:
Optimality and beyond: The case of 4×4 s-boxes.
2014 IEEE International Symposium on Hardware-Oriented Security and Trust (HOST),
80--83
(2014)
\doiurl{10.1109/HST.2014.6855573}
\end{botherref}
\endbibitem

\bibitem[\protect\citeauthoryear{Jana and Paul}{2022}]{Jana}
\begin{botherref}
\oauthor{\bsnm{Jana}, \binits{A.}},
\oauthor{\bsnm{Paul}, \binits{G.}}:
Differential fault attack on photon-beetle.
Proceedings of the 2022 Workshop on Attacks and Solutions in Hardware Security,
25--34
(2022)
\doiurl{10.1145/3560834.3563824}
\end{botherref}
\endbibitem

\bibitem[\protect\citeauthoryear{Dobraunig and Mennink}{2020}]{dobraunigkey}
\begin{botherref}
\oauthor{\bsnm{Dobraunig}, \binits{C.}},
\oauthor{\bsnm{Mennink}, \binits{B.}}:
Key recovery attack on photon-beetle.
NIST, Gaithersburg, MD, USA, Tech. Rep
(2020)
\end{botherref}
\endbibitem

\bibitem[\protect\citeauthoryear{Inoue et~al.}{2022}]{Akiko}
\begin{botherref}
\oauthor{\bsnm{Inoue}, \binits{A.}},
\oauthor{\bsnm{Iwata}, \binits{T.}},
\oauthor{\bsnm{Minematsu}, \binits{K.}}:
Analyzing the provable security bounds of gift-cofb and photon-beetle.
Cryptology ePrint Archive, Paper 2022/001
(2022).
\url{https://eprint.iacr.org/2022/001}
\end{botherref}
\endbibitem

\bibitem[\protect\citeauthoryear{Guo et~al.}{2011}]{JGuo}
\begin{botherref}
\oauthor{\bsnm{Guo}, \binits{J.}},
\oauthor{\bsnm{Peyrin}, \binits{T.}},
\oauthor{\bsnm{Poschmann}, \binits{A.}}:
The photon family of lightweight hash functions.
Advances in Cryptology -- CRYPTO 2011,
222--239
(2011)
\end{botherref}
\endbibitem

\bibitem[\protect\citeauthoryear{Cui et~al.}{2017}]{cuitinting}
\begin{botherref}
\oauthor{\bsnm{Cui}, \binits{T.}},
\oauthor{\bsnm{Sun}, \binits{L.}},
\oauthor{\bsnm{Chen}, \binits{H.}},
\oauthor{\bsnm{Wang}, \binits{M.}}:
Statistical integral distinguisher with multi-structure and its application on aes.
Information Security and Privacy,
402--420
(2017)
\end{botherref}
\endbibitem

\bibitem[\protect\citeauthoryear{Jean et~al.}{2014}]{Jean}
\begin{botherref}
\oauthor{\bsnm{Jean}, \binits{J.}},
\oauthor{\bsnm{Naya-Plasencia}, \binits{M.}},
\oauthor{\bsnm{Peyrin}, \binits{T.}}:
Multiple limited-birthday distinguishers and applications.
Selected Areas in Cryptography -- SAC 2013,
533--550
(2014)
\end{botherref}
\endbibitem

\bibitem[\protect\citeauthoryear{Zhou et~al.}{2021}]{zhou2021interpolation}
\begin{barticle}
\bauthor{\bsnm{Zhou}, \binits{H.}},
\bauthor{\bsnm{Zong}, \binits{R.}},
\bauthor{\bsnm{Dong}, \binits{X.}},
\bauthor{\bsnm{Jia}, \binits{K.}},
\bauthor{\bsnm{Meier}, \binits{W.}}:
\batitle{Interpolation attacks on round-reduced elephant, kravatte and xoofff}.
\bjtitle{The Computer Journal}
\bvolume{64}(\bissue{4}),
\bfpage{628}--\blpage{638}
(\byear{2021})
\end{barticle}
\endbibitem

\bibitem[\protect\citeauthoryear{Joshi and Mazumdar}{2021}]{Priyanka}
\begin{botherref}
\oauthor{\bsnm{Joshi}, \binits{P.}},
\oauthor{\bsnm{Mazumdar}, \binits{B.}}:
Single event transient fault analysis of {ELEPHANT} cipher.
CoRR
\textbf{abs/2106.09536}
(2021)
{\href{https://arxiv.org/abs/2106.09536}{{2106.09536}}}
\end{botherref}
\endbibitem

\bibitem[\protect\citeauthoryear{Meraneh et~al.}{2022}]{meraneh2022blind}
\begin{botherref}
\oauthor{\bsnm{Meraneh}, \binits{A.H.}},
\oauthor{\bsnm{Clavier}, \binits{C.}},
\oauthor{\bsnm{Le~Bouder}, \binits{H.}},
\oauthor{\bsnm{Maillard}, \binits{J.}},
\oauthor{\bsnm{Thomas}, \binits{G.}}:
Blind side channel on the elephant lfsr.
SECRYPT 2022
(2022)
\end{botherref}
\endbibitem

\bibitem[\protect\citeauthoryear{Beyne et~al.}{2022}]{Beyne}
\begin{botherref}
\oauthor{\bsnm{Beyne}, \binits{T.}},
\oauthor{\bsnm{Chen}, \binits{Y.L.}},
\oauthor{\bsnm{Dobraunig}, \binits{C.}},
\oauthor{\bsnm{Mennink}, \binits{B.}}:
Multi-user security of the elephant v2 authenticated encryption mode.
Selected Areas in Cryptography,
155--178
(2022)
\end{botherref}
\endbibitem

\bibitem[\protect\citeauthoryear{Alagic et~al.}{2022}]{Gorjan}
\begin{botherref}
\oauthor{\bsnm{Alagic}, \binits{G.}},
\oauthor{\bsnm{Bai}, \binits{C.}},
\oauthor{\bsnm{Katz}, \binits{J.}},
\oauthor{\bsnm{Majenz}, \binits{C.}},
\oauthor{\bsnm{Struck}, \binits{P.}}:
Post-quantum security of the (tweakable) fx construction, and applications.
Cryptology ePrint Archive, Paper 2022/1097
(2022).
\url{https://eprint.iacr.org/2022/1097}
\end{botherref}
\endbibitem

\bibitem[\protect\citeauthoryear{Tolba et~al.}{2017}]{Tolba}
\begin{botherref}
\oauthor{\bsnm{Tolba}, \binits{M.}},
\oauthor{\bsnm{Abdelkhalek}, \binits{A.}},
\oauthor{\bsnm{Youssef}, \binits{A.M.}}:
Impossible differential cryptanalysis of reduced-round skinny.
Progress in Cryptology - AFRICACRYPT 2017,
117--134
(2017)
\end{botherref}
\endbibitem

\bibitem[\protect\citeauthoryear{Habu et~al.}{2022}]{rom1}
\begin{botherref}
\oauthor{\bsnm{Habu}, \binits{M.}},
\oauthor{},
\oauthor{\bsnm{Minematsu}, \binits{K.}},
\oauthor{\bsnm{Iwata}, \binits{T.}}:
Matching attacks on romulus-m.
Cryptology ePrint Archive, Report 2022/369
(2022)
\end{botherref}
\endbibitem

\bibitem[\protect\citeauthoryear{Dong et~al.}{2021}]{rom2}
\begin{botherref}
\oauthor{\bsnm{Dong}, \binits{X.}},
\oauthor{\bsnm{Hua}, \binits{J.}},
\oauthor{\bsnm{Sun}, \binits{S.}},
\oauthor{\bsnm{Li}, \binits{Z.}},
\oauthor{\bsnm{Wang}, \binits{X.}},
\oauthor{\bsnm{Hu}, \binits{L.}}:
Meet-in-the-middle attacks revisited: Key-recovery, collision, and preimage attacks.
Cryptology ePrint Archive, Report 2021/427
(2021)
\end{botherref}
\endbibitem

\bibitem[\protect\citeauthoryear{Hadipour et~al.}{2022}]{Hosein}
\begin{barticle}
\bauthor{\bsnm{Hadipour}, \binits{H.}},
\bauthor{\bsnm{Sadeghi}, \binits{S.}},
\bauthor{\bsnm{Eichlseder}, \binits{M.}}:
\batitle{Finding the impossible: Automated search for full impossible-differential, zero-correlation, and integral attacks}.
\bjtitle{Cryptology ePrint Archive, Paper 2022/1147}
(\byear{2022})
\doiurl{10.1007/978-3-031-30634-1_5} .
\bcomment{\url{https://eprint.iacr.org/2022/1147}}
\end{barticle}
\endbibitem

\bibitem[\protect\citeauthoryear{Shi et~al.}{2018}]{Shi}
\begin{botherref}
\oauthor{\bsnm{Shi}, \binits{D.}},
\oauthor{\bsnm{Sun}, \binits{S.}},
\oauthor{\bsnm{Derbez}, \binits{P.}},
\oauthor{\bsnm{Todo}, \binits{Y.}},
\oauthor{\bsnm{Sun}, \binits{B.}},
\oauthor{\bsnm{Hu}, \binits{L.}}:
Programming the demirci-sel{\c{c}}uk meet-in-the-middle attack with constraints.
Advances in Cryptology -- ASIACRYPT 2018,
3--34
(2018)
\end{botherref}
\endbibitem

\bibitem[\protect\citeauthoryear{Qin et~al.}{2021}]{Qin_Dong}
\begin{barticle}
\bauthor{\bsnm{Qin}, \binits{L.}},
\bauthor{\bsnm{Dong}, \binits{X.}},
\bauthor{\bsnm{Wang}, \binits{X.}},
\bauthor{\bsnm{Jia}, \binits{K.}},
\bauthor{\bsnm{Liu}, \binits{Y.}}:
\batitle{Automated search oriented to key recovery on ciphers with linear key schedule: Applications to boomerangs in skinny and forkskinny}.
\bjtitle{IACR Transactions on Symmetric Cryptology}
\bvolume{2021}(\bissue{2}),
\bfpage{249}--\blpage{291}
(\byear{2021})
\doiurl{10.46586/tosc.v2021.i2.249-291}
\end{barticle}
\endbibitem

\bibitem[\protect\citeauthoryear{Bijwe et~al.}{2022}]{Bijwe}
\begin{botherref}
\oauthor{\bsnm{Bijwe}, \binits{S.}},
\oauthor{\bsnm{Chauhan}, \binits{A.K.}},
\oauthor{\bsnm{Sanadhya}, \binits{S.K.}}:
Implementing grover oracle for lightweight block ciphers under depth constraints.
Information Security and Privacy,
85--105
(2022)
\end{botherref}
\endbibitem

\bibitem[\protect\citeauthoryear{Tezcan}{2016}]{cihangir2}
\begin{botherref}
\oauthor{\bsnm{Tezcan}}:
Truncated, impossible, and improbable differential analysis of ascon.
Cryptology ePrint Archive, Report 2016/490
(2016)
\end{botherref}
\endbibitem

\bibitem[\protect\citeauthoryear{Peng et~al.}{2025}]{peng2025improved}
\begin{bchapter}
\bauthor{\bsnm{Peng}, \binits{S.}},
\bauthor{\bsnm{Hu}, \binits{K.}},
\bauthor{\bsnm{He}, \binits{J.}},
\bauthor{\bsnm{Wang}, \binits{M.}}:
\bctitle{Improved key recovery attacks of ascon}.
In: \bbtitle{Cryptographers’ Track at the RSA Conference},
pp. \bfpage{123}--\blpage{146}
(\byear{2025}).
\bcomment{Springer}
\end{bchapter}
\endbibitem

\bibitem[\protect\citeauthoryear{Nguyen et~al.}{2025}]{nguyencorrelation}
\begin{botherref}
\oauthor{\bsnm{Nguyen}, \binits{V.S.}},
\oauthor{\bsnm{Grosso}, \binits{V.}},
\oauthor{\bsnm{Cayrel}, \binits{P.-L.}}:
Correlation power analysis on ascon with multi-bit selection function.
International Conference on Security and Cryptography (SECRYPT)
(2025)
\end{botherref}
\endbibitem

\bibitem[\protect\citeauthoryear{Jana}{2024}]{jana2024differentialascon}
\begin{botherref}
\oauthor{\bsnm{Jana}, \binits{A.}}:
Differential fault attack on ascon cipher.
International Conference on Cryptology in India,
53--72
(2024).
Springer
\end{botherref}
\endbibitem

\bibitem[\protect\citeauthoryear{Zhai et~al.}{2024}]{zhai2024improved}
\begin{botherref}
\oauthor{\bsnm{Zhai}, \binits{D.}},
\oauthor{\bsnm{Bai}, \binits{W.}},
\oauthor{\bsnm{Fu}, \binits{J.}},
\oauthor{\bsnm{Gao}, \binits{H.}},
\oauthor{\bsnm{Zhu}, \binits{X.}}:
Improved 2-round collision attack on iot hash standard ascon-hash.
Heliyon
\textbf{10}(5)
(2024)
\end{botherref}
\endbibitem

\bibitem[\protect\citeauthoryear{Fu et~al.}{2025}]{fu2025preimage}
\begin{barticle}
\bauthor{\bsnm{Fu}, \binits{Q.}},
\bauthor{\bsnm{Luo}, \binits{Y.}},
\bauthor{\bsnm{Yang}, \binits{Q.}},
\bauthor{\bsnm{Song}, \binits{L.}}:
\batitle{Preimage and collision attacks on reduced ascon using algebraic strategies}.
\bjtitle{Cybersecurity}
\bvolume{8}(\bissue{1}),
\bfpage{1}--\blpage{17}
(\byear{2025})
\end{barticle}
\endbibitem

\bibitem[\protect\citeauthoryear{Zhu et~al.}{2019}]{Baoyu}
\begin{botherref}
\oauthor{\bsnm{Zhu}, \binits{B.}},
\oauthor{\bsnm{Dong}, \binits{X.}},
\oauthor{\bsnm{Yu}, \binits{H.}}:
Milp-based differential attack on round-reduced gift.
Topics in Cryptology -- CT-RSA 2019,
372--390
(2019)
\end{botherref}
\endbibitem

\bibitem[\protect\citeauthoryear{Jana and Paul}{2024}]{jana2024differential}
\begin{barticle}
\bauthor{\bsnm{Jana}, \binits{A.}},
\bauthor{\bsnm{Paul}, \binits{G.}}:
\batitle{Differential fault attack on spn-based sponge and siv-like ae schemes}.
\bjtitle{Journal of Cryptographic Engineering}
\bvolume{14}(\bissue{2}),
\bfpage{363}--\blpage{381}
(\byear{2024})
\end{barticle}
\endbibitem

\bibitem[\protect\citeauthoryear{Krämer et~al.}{2024}]{Krämer_Struck_Weishäupl_2024}
\begin{barticle}
\bauthor{\bsnm{Krämer}, \binits{J.}},
\bauthor{\bsnm{Struck}, \binits{P.}},
\bauthor{\bsnm{Weishäupl}, \binits{M.}}:
\batitle{Committing ae from sponges: Security analysis of the nist lwc finalists}.
\bjtitle{IACR Transactions on Symmetric Cryptology}
\bvolume{2024}(\bissue{4}),
\bfpage{191}--\blpage{248}
(\byear{2024})
\doiurl{10.46586/tosc.v2024.i4.191-248}
\end{barticle}
\endbibitem

\bibitem[\protect\citeauthoryear{Dong et~al.}{2024}]{dong2024generic}
\begin{botherref}
\oauthor{\bsnm{Dong}, \binits{X.}},
\oauthor{\bsnm{Zhao}, \binits{B.}},
\oauthor{\bsnm{Qin}, \binits{L.}},
\oauthor{\bsnm{Hou}, \binits{Q.}},
\oauthor{\bsnm{Zhang}, \binits{S.}},
\oauthor{\bsnm{Wang}, \binits{X.}}:
Generic mitm attack frameworks on sponge constructions.
Annual International Cryptology Conference,
3--37
(2024).
Springer
\end{botherref}
\endbibitem

\bibitem[\protect\citeauthoryear{Zhang et~al.}{2024}]{Zhang_Wang_Tang_2024}
\begin{barticle}
\bauthor{\bsnm{Zhang}, \binits{J.}},
\bauthor{\bsnm{Wang}, \binits{H.}},
\bauthor{\bsnm{Tang}, \binits{D.}}:
\batitle{Impossible boomerang attacks revisited: Applications to deoxys-bc, joltik-bc and skinny}.
\bjtitle{IACR Transactions on Symmetric Cryptology}
\bvolume{2024}(\bissue{2}),
\bfpage{254}--\blpage{295}
(\byear{2024})
\doiurl{10.46586/tosc.v2024.i2.254-295}
\end{barticle}
\endbibitem

\bibitem[\protect\citeauthoryear{Ahmadian et~al.}{2024}]{10.1007/978-3-031-58716-0_10}
\begin{botherref}
\oauthor{\bsnm{Ahmadian}, \binits{Z.}},
\oauthor{\bsnm{Khalesi}, \binits{A.}},
\oauthor{\bsnm{M'Foukh}, \binits{D.}},
\oauthor{\bsnm{Moghimi}, \binits{H.}},
\oauthor{\bsnm{Naya-Plasencia}, \binits{M.}}:
Improved differential meet-in-the-middle cryptanalysis.
Advances in Cryptology -- EUROCRYPT 2024,
280--309
(2024)
\end{botherref}
\endbibitem

\bibitem[\protect\citeauthoryear{Song et~al.}{2025}]{song2025generalized}
\begin{botherref}
\oauthor{\bsnm{Song}, \binits{L.}},
\oauthor{\bsnm{Fu}, \binits{Q.}},
\oauthor{\bsnm{Yang}, \binits{Q.}},
\oauthor{\bsnm{Lv}, \binits{Y.}},
\oauthor{\bsnm{Hu}, \binits{L.}}:
Generalized impossible differential attacks on block ciphers: application to skinny and forkskinny.
Designs, Codes and Cryptography,
1--45
(2025)
\end{botherref}
\endbibitem

\bibitem[\protect\citeauthoryear{Perrin}{2013}]{perrin2013properties}
\begin{botherref}
\oauthor{\bsnm{Perrin}, \binits{L.P.}}:
On the properties of s-boxes: A study of differentially 6-uniform monomials over finite fields of characteristic 2
(2013)
\end{botherref}
\endbibitem

\bibitem[\protect\citeauthoryear{Heys}{2002}]{heys2002tutorial}
\begin{barticle}
\bauthor{\bsnm{Heys}, \binits{H.M.}}:
\batitle{A tutorial on linear and differential cryptanalysis}.
\bjtitle{Cryptologia}
\bvolume{26}(\bissue{3}),
\bfpage{189}--\blpage{221}
(\byear{2002})
\end{barticle}
\endbibitem

\bibitem[\protect\citeauthoryear{Ishfaq}{2018}]{ishfaq2018matlab}
\begin{botherref}
\oauthor{\bsnm{Ishfaq}, \binits{F.}}:
A matlab tool for the analysis of cryptographic properties of s-boxes.
Capital University
(2018)
\end{botherref}
\endbibitem

\bibitem[\protect\citeauthoryear{Seitkulov et~al.}{2021}]{seitkulov2021}
\begin{barticle}
\bauthor{\bsnm{Seitkulov}, \binits{Y.}},
\bauthor{\bsnm{Ospanov}, \binits{R.}},
\bauthor{\bsnm{Yergaliyeva}, \binits{B.}}:
\batitle{On cryptographic properties of s-boxes}.
\bjtitle{Engineering Journal of Satbayev University}
\bvolume{143}(\bissue{4}),
\bfpage{96}--\blpage{103}
(\byear{2021})
\end{barticle}
\endbibitem

\bibitem[\protect\citeauthoryear{Nitaj et~al.}{2020}]{nitaj2020new}
\begin{barticle}
\bauthor{\bsnm{Nitaj}, \binits{A.}},
\bauthor{\bsnm{Susilo}, \binits{W.}},
\bauthor{\bsnm{Tonien}, \binits{J.}}:
\batitle{A new improved aes s-box with enhanced properties}.
\bjtitle{Cryptology ePrint Archive, Paper 2020/1597}
(\byear{2020})
\doiurl{10.1007/978-3-030-55304-3}
\end{barticle}
\endbibitem

\bibitem[\protect\citeauthoryear{afify* et~al.}{2020}]{abeer}
\begin{barticle}
\bauthor{\bsnm{afify*}, \binits{E.w.}},
\bauthor{\bsnm{sobky}, \binits{W.I.}},
\bauthor{\bsnm{Twakol}, \binits{A.}},
\bauthor{\bsnm{Alez}, \binits{R.A.}}:
\batitle{Algebraic construction of powerful substitution box}.
\bjtitle{International Journal of Recent Technology and Engineering (IJRTE)}
\bvolume{8}(\bissue{6}),
\bfpage{405}--\blpage{409}
(\byear{2020})
\doiurl{10.35940/ijrte.d8279.038620}
\end{barticle}
\endbibitem

\bibitem[\protect\citeauthoryear{Zahid et~al.}{2021}]{zahid2021dynamic}
\begin{barticle}
\bauthor{\bsnm{Zahid}, \binits{A.H.}},
\bauthor{\bsnm{Rashid}, \binits{H.}},
\bauthor{\bsnm{Shaban}, \binits{M.M.U.}},
\bauthor{\bsnm{Ahmad}, \binits{S.}},
\bauthor{\bsnm{Ahmed}, \binits{E.}},
\bauthor{\bsnm{Amjad}, \binits{M.T.}},
\bauthor{\bsnm{Baig}, \binits{M.A.T.}},
\bauthor{\bsnm{Arshad}, \binits{M.J.}},
\bauthor{\bsnm{Tariq}, \binits{M.N.}},
\bauthor{\bsnm{Tariq}, \binits{M.W.}}, \betal:
\batitle{Dynamic s-box design using a novel square polynomial transformation and permutation}.
\bjtitle{IEEE Access}
\bvolume{9},
\bfpage{82390}--\blpage{82401}
(\byear{2021})
\end{barticle}
\endbibitem

\bibitem[\protect\citeauthoryear{Evertse}{1987}]{evertse1987linear}
\begin{botherref}
\oauthor{\bsnm{Evertse}, \binits{J.-H.}}:
Linear structures in blockciphers.
Workshop on the Theory and Application of of Cryptographic Techniques,
249--266
(1987).
Springer
\end{botherref}
\endbibitem

\bibitem[\protect\citeauthoryear{Nyberg}{2023}]{nyberg2023modifications}
\begin{barticle}
\bauthor{\bsnm{Nyberg}, \binits{K.}}:
\batitle{Modifications of bijective s-boxes with linear structures}.
\bjtitle{Cryptography and Communications}
\bvolume{15}(\bissue{3}),
\bfpage{617}--\blpage{625}
(\byear{2023})
\end{barticle}
\endbibitem

\bibitem[\protect\citeauthoryear{Boura and Canteaut}{2018}]{boura2018boomerang}
\begin{botherref}
\oauthor{\bsnm{Boura}, \binits{C.}},
\oauthor{\bsnm{Canteaut}, \binits{A.}}:
On the boomerang uniformity of cryptographic sboxes.
IACR Transactions on Symmetric Cryptology,
290--310
(2018)
\end{botherref}
\endbibitem

\bibitem[\protect\citeauthoryear{Tezcan and {\"O}zbudak}{2014}]{tezcan2014differential}
\begin{botherref}
\oauthor{\bsnm{Tezcan}, \binits{C.}},
\oauthor{\bsnm{{\"O}zbudak}, \binits{F.}}:
Differential factors: Improved attacks on serpent.
International Workshop on Lightweight Cryptography for Security and Privacy,
69--84
(2014).
Springer
\end{botherref}
\endbibitem

\bibitem[\protect\citeauthoryear{Bar-On et~al.}{2019}]{bar2019dlct}
\begin{botherref}
\oauthor{\bsnm{Bar-On}, \binits{A.}},
\oauthor{\bsnm{Dunkelman}, \binits{O.}},
\oauthor{\bsnm{Keller}, \binits{N.}},
\oauthor{\bsnm{Weizman}, \binits{A.}}:
Dlct: a new tool for differential-linear cryptanalysis.
Advances in Cryptology--EUROCRYPT 2019: 38th Annual International Conference on the Theory and Applications of Cryptographic Techniques, Darmstadt, Germany, May 19--23, 2019, Proceedings, Part I 38,
313--342
(2019).
Springer
\end{botherref}
\endbibitem

\bibitem[\protect\citeauthoryear{Kim et~al.}{2018}]{kim2018improved}
\begin{barticle}
\bauthor{\bsnm{Kim}, \binits{H.}},
\bauthor{\bsnm{Kim}, \binits{S.}},
\bauthor{\bsnm{Hong}, \binits{D.}},
\bauthor{\bsnm{Sung}, \binits{J.}},
\bauthor{\bsnm{Hong}, \binits{S.}}:
\batitle{Improved differential-linear cryptanalysis using dlct}.
\bjtitle{Journal of The Korea Institute of Information Security \& Cryptology}
\bvolume{28}(\bissue{6}),
\bfpage{1379}--\blpage{1392}
(\byear{2018})
\end{barticle}
\endbibitem

\bibitem[\protect\citeauthoryear{Canteaut}{2016}]{canteaut2016lecture}
\begin{botherref}
\oauthor{\bsnm{Canteaut}, \binits{A.}}:
Lecture notes on cryptographic boolean functions.
Inria, Paris, France
\textbf{3}
(2016)
\end{botherref}
\endbibitem

\bibitem[\protect\citeauthoryear{Guilley et~al.}{2004}]{guilley2004differential}
\begin{botherref}
\oauthor{\bsnm{Guilley}, \binits{S.}},
\oauthor{\bsnm{Hoogvorst}, \binits{P.}},
\oauthor{\bsnm{Pacalet}, \binits{R.}}:
Differential power analysis model and some results.
CARDIS,
127--142
(2004)
\end{botherref}
\endbibitem

\bibitem[\protect\citeauthoryear{Prouff}{2005}]{prouff2005dpa}
\begin{botherref}
\oauthor{\bsnm{Prouff}, \binits{E.}}:
Dpa attacks and s-boxes.
International Workshop on Fast Software Encryption,
424--441
(2005).
Springer
\end{botherref}
\endbibitem

\bibitem[\protect\citeauthoryear{Naseer et~al.}{2022}]{9990069}
\begin{botherref}
\oauthor{\bsnm{Naseer}, \binits{M.}},
\oauthor{\bsnm{Tariq}, \binits{S.}},
\oauthor{\bsnm{Riaz}, \binits{N.}}:
Substitution layer analysis of nist lightweight cryptography competition finalists.
2022 19th International Bhurban Conference on Applied Sciences and Technology (IBCAST),
659--664
(2022)
\doiurl{10.1109/IBCAST54850.2022.9990069}
\end{botherref}
\endbibitem

\bibitem[\protect\citeauthoryear{Bao et~al.}{2019}]{peigen}
\begin{botherref}
\oauthor{\bsnm{Bao}, \binits{Z.}},
\oauthor{\bsnm{Guo}, \binits{J.}},
\oauthor{\bsnm{Ling}, \binits{S.}},
\oauthor{\bsnm{Sasaki}, \binits{Y.}}:
{SoK}: Peigen -- a platform for evaluation, implementation, and generation of s-boxes
(2019)
\end{botherref}
\endbibitem

\bibitem[\protect\citeauthoryear{Dobraunig et~al.}{2021}]{dobraunig2021ascon}
\begin{barticle}
\bauthor{\bsnm{Dobraunig}, \binits{C.}},
\bauthor{\bsnm{Eichlseder}, \binits{M.}},
\bauthor{\bsnm{Mendel}, \binits{F.}},
\bauthor{\bsnm{Schl{\"a}ffer}, \binits{M.}}:
\batitle{Ascon v1. 2: Lightweight authenticated encryption and hashing}.
\bjtitle{Journal of Cryptology}
\bvolume{34}(\bissue{3}),
\bfpage{1}--\blpage{42}
(\byear{2021})
\end{barticle}
\endbibitem

\bibitem[\protect\citeauthoryear{Dobraunig et~al.}{2020}]{dobraunig2020isap}
\begin{botherref}
\oauthor{\bsnm{Dobraunig}, \binits{C.}},
\oauthor{\bsnm{Eichlseder}, \binits{M.}},
\oauthor{\bsnm{Mangard}, \binits{S.}},
\oauthor{\bsnm{Mendel}, \binits{F.}},
\oauthor{\bsnm{Mennink}, \binits{B.}},
\oauthor{\bsnm{Primas}, \binits{R.}},
\oauthor{\bsnm{Unterluggauer}, \binits{T.}}:
Isap v2.0.
NIST Lightweight Cryptography
(2020)
\end{botherref}
\endbibitem

\bibitem[\protect\citeauthoryear{Banik et~al.}{2019}]{cryptoeprint:2020:738}
\begin{botherref}
\oauthor{\bsnm{Banik}, \binits{S.}},
\oauthor{\bsnm{Chakraborti}, \binits{A.}},
\oauthor{\bsnm{Iwata}, \binits{T.}},
\oauthor{\bsnm{Minematsu}, \binits{K.}},
\oauthor{\bsnm{Nandi}, \binits{M.}},
\oauthor{\bsnm{Peyrin}, \binits{T.}},
\oauthor{\bsnm{Sasaki}, \binits{Y.}},
\oauthor{\bsnm{Todo}, \binits{Y.}}:
Gift-cofb.
NIST
(2019)
\end{botherref}
\endbibitem

\bibitem[\protect\citeauthoryear{Bao et~al.}{2019}]{bao2019photon}
\begin{barticle}
\bauthor{\bsnm{Bao}, \binits{Z.}},
\bauthor{\bsnm{Chakraborti}, \binits{A.}},
\bauthor{\bsnm{Datta}, \binits{N.}},
\bauthor{\bsnm{Guo}, \binits{J.}},
\bauthor{\bsnm{Nandi}, \binits{M.}},
\bauthor{\bsnm{Peyrin}, \binits{T.}},
\bauthor{\bsnm{Yasuda}, \binits{K.}}:
\batitle{Photon-beetle authenticated encryption and hash family}.
\bjtitle{NIST Lightweight Compet. Round}
\bvolume{1},
\bfpage{115}
(\byear{2019})
\end{barticle}
\endbibitem

\bibitem[\protect\citeauthoryear{Beyne et~al.}{2021}]{mennink2021elephant}
\begin{botherref}
\oauthor{\bsnm{Beyne}, \binits{T.}},
\oauthor{\bsnm{Chen}, \binits{Y.L.}},
\oauthor{\bsnm{Dobraunig}, \binits{C.}},
\oauthor{\bsnm{Mennink}, \binits{B.}}:
Elephant v2.
NIST Lightweight Cryptography
(2021)
\end{botherref}
\endbibitem

\bibitem[\protect\citeauthoryear{Guo et~al.}{2021}]{guodesigners}
\begin{botherref}
\oauthor{\bsnm{Guo}, \binits{C.}},
\oauthor{\bsnm{Iwata}, \binits{T.}},
\oauthor{\bsnm{Khairallah}, \binits{M.}},
\oauthor{\bsnm{Minematsu}, \binits{K.}},
\oauthor{\bsnm{Peyrin}, \binits{T.}}:
Romulus
(2021)
\end{botherref}
\endbibitem

\bibitem[\protect\citeauthoryear{Beierle et~al.}{2016}]{Skinny}
\begin{botherref}
\oauthor{\bsnm{Beierle}, \binits{C.}},
\oauthor{\bsnm{Jean}, \binits{J.}},
\oauthor{\bsnm{K{\"o}lbl}, \binits{S.}},
\oauthor{\bsnm{Leander}, \binits{G.}},
\oauthor{\bsnm{Moradi}, \binits{A.}},
\oauthor{\bsnm{Peyrin}, \binits{T.}},
\oauthor{\bsnm{Sasaki}, \binits{Y.}},
\oauthor{\bsnm{Sasdrich}, \binits{P.}},
\oauthor{\bsnm{Sim}, \binits{S.M.}}:
The skinny family of block ciphers and its low-latency variant mantis.
Advances in Cryptology--CRYPTO 2016: 36th Annual International Cryptology Conference, Santa Barbara, CA, USA, August 14-18, 2016, Proceedings, Part II 36,
123--153
(2016).
Springer
\end{botherref}
\endbibitem

\bibitem[\protect\citeauthoryear{Cubero}{2015}]{joscubero}
\begin{botherref}
\oauthor{\bsnm{Cubero}, \binits{J.A.{\'A}.}}:
Vector boolean functions: Applications in symmetric cryptography
(2015)
\doiurl{10.13140/RG.2.2.12540.23685}
\end{botherref}
\endbibitem

\bibitem[\protect\citeauthoryear{Khadem and Rajavzade}{2021}]{khadem2021construction}
\begin{botherref}
\oauthor{\bsnm{Khadem}, \binits{B.}},
\oauthor{\bsnm{Rajavzade}, \binits{S.}}:
Construction of side channel attacks resistant s-boxes using genetic algorithms based on coordinate functions.
arXiv preprint arXiv:2102.09799
(2021)
\end{botherref}
\endbibitem

\bibitem[\protect\citeauthoryear{O'Connor}{1995}]{10.1007/3-540-60590-8_10}
\begin{botherref}
\oauthor{\bsnm{O'Connor}, \binits{L.}}:
Properties of linear approximation tables.
Fast Software Encryption (FSE),
131--136
(1995)
\end{botherref}
\endbibitem

\bibitem[\protect\citeauthoryear{Bakunina and Dykyi}{2022}]{bakunina2022synthesis}
\begin{botherref}
\oauthor{\bsnm{Bakunina}, \binits{E.}},
\oauthor{\bsnm{Dykyi}, \binits{O.}}:
Synthesis method for s-boxes satisfying the criterion of correlation immunity of boolean and 4-functions.
Journal of Discrete Mathematical Sciences and Cryptography,
1--13
(2022)
\end{botherref}
\endbibitem

\bibitem[\protect\citeauthoryear{{\'A}lvarez-Cubero and Zufiria}{2012}]{alvarez2012cryptographic}
\begin{bbook}
\bauthor{\bsnm{{\'A}lvarez-Cubero}, \binits{J.A.}},
\bauthor{\bsnm{Zufiria}, \binits{P.J.}}:
\bbtitle{Cryptographic Criteria on Vector Boolean Functions}.
\bpublisher{InTech}, \blocation{???}
(\byear{2012})
\end{bbook}
\endbibitem

\bibitem[\protect\citeauthoryear{Carlet}{2021}]{carlet2021boolean}
\begin{bbook}
\bauthor{\bsnm{Carlet}, \binits{C.}}:
\bbtitle{Boolean Functions for Cryptography and Coding Theory}.
\bpublisher{Cambridge University Press}, \blocation{???}
(\byear{2021})
\end{bbook}
\endbibitem

\bibitem[\protect\citeauthoryear{Sarkar and Syed}{2017}]{sarkar2017bounds}
\begin{botherref}
\oauthor{\bsnm{Sarkar}, \binits{S.}},
\oauthor{\bsnm{Syed}, \binits{H.}}:
Bounds on differential and linear branch number of permutations.
Cryptology ePrint Archive
(2017)
\end{botherref}
\endbibitem

\bibitem[\protect\citeauthoryear{Song et~al.}{2019}]{song2019boomerang}
\begin{botherref}
\oauthor{\bsnm{Song}, \binits{L.}},
\oauthor{\bsnm{Qin}, \binits{X.}},
\oauthor{\bsnm{Hu}, \binits{L.}}:
Boomerang connectivity table revisited. application to skinny and aes.
IACR Transactions on Symmetric Cryptology,
118--141
(2019)
\end{botherref}
\endbibitem

\bibitem[\protect\citeauthoryear{Canteaut et~al.}{2021}]{canteaut2021autocorrelations}
\begin{botherref}
\oauthor{\bsnm{Canteaut}, \binits{A.}},
\oauthor{\bsnm{K{\"o}lsch}, \binits{L.}},
\oauthor{\bsnm{Li}, \binits{C.}},
\oauthor{\bsnm{Li}, \binits{C.}},
\oauthor{\bsnm{Li}, \binits{K.}},
\oauthor{\bsnm{Qu}, \binits{L.}},
\oauthor{\bsnm{Wiemer}, \binits{F.}}:
Autocorrelations of vectorial boolean functions.
International Conference on Cryptology and Information Security in Latin America,
233--253
(2021).
Springer
\end{botherref}
\endbibitem

\bibitem[\protect\citeauthoryear{Canteaut et~al.}{2019}]{canteaut2019differential}
\begin{botherref}
\oauthor{\bsnm{Canteaut}, \binits{A.}},
\oauthor{\bsnm{Kolsch}, \binits{L.}},
\oauthor{\bsnm{Li}, \binits{C.}},
\oauthor{\bsnm{Li}, \binits{C.}},
\oauthor{\bsnm{Li}, \binits{K.}},
\oauthor{\bsnm{Qu}, \binits{L.}},
\oauthor{\bsnm{Wiemer}, \binits{F.}}:
On the differential-linear connectivity table of vectorial boolean functions.
arXiv preprint arXiv:1908.07445
(2019)
\end{botherref}
\endbibitem

\bibitem[\protect\citeauthoryear{Farwa et~al.}{2016}]{farwa2016highly}
\begin{barticle}
\bauthor{\bsnm{Farwa}, \binits{S.}},
\bauthor{\bsnm{Shah}, \binits{T.}},
\bauthor{\bsnm{Idrees}, \binits{L.}}:
\batitle{A highly nonlinear s-box based on a fractional linear transformation}.
\bjtitle{Springer}
\bvolume{5}(\bissue{1}),
\bfpage{1}--\blpage{12}
(\byear{2016})
\end{barticle}
\endbibitem

\bibitem[\protect\citeauthoryear{Sagd{\i}{\c{c}}oglu}{2003}]{sagdiccoglu2003cryptological}
\begin{botherref}
\oauthor{\bsnm{Sagd{\i}{\c{c}}oglu}, \binits{S.}}:
Cryptological viewpoint of boolean functions
(2003)
\end{botherref}
\endbibitem

\bibitem[\protect\citeauthoryear{Picek et~al.}{2016}]{picek2016search}
\begin{botherref}
\oauthor{\bsnm{Picek}, \binits{S.}},
\oauthor{\bsnm{Yang}, \binits{B.}},
\oauthor{\bsnm{Mentens}, \binits{N.}}:
A search strategy to optimize the affine variant properties of s-boxes.
International Workshop on the Arithmetic of Finite Fields,
208--223
(2016).
Springer
\end{botherref}
\endbibitem

\bibitem[\protect\citeauthoryear{Musukwa et~al.}{2019}]{musukwa2019some}
\begin{botherref}
\oauthor{\bsnm{Musukwa}, \binits{A.}},
\oauthor{\bsnm{Sala}, \binits{M.}},
\oauthor{\bsnm{Zaninelli}, \binits{M.}}:
On some cryptographic properties of boolean functions and their second-order derivatives.
arXiv preprint arXiv:1909.10586
(2019)
\end{botherref}
\endbibitem

\bibitem[\protect\citeauthoryear{Cui et~al.}{2011}]{cui2011improved}
\begin{barticle}
\bauthor{\bsnm{Cui}, \binits{J.}},
\bauthor{\bsnm{Huang}, \binits{L.}},
\bauthor{\bsnm{Zhong}, \binits{H.}},
\bauthor{\bsnm{Chang}, \binits{C.}},
\bauthor{\bsnm{Yang}, \binits{W.}}:
\batitle{An improved aes s-box and its performance analysis}.
\bjtitle{International Journal of Innovative Computing, Information and Control}
\bvolume{7}(\bissue{5}),
\bfpage{2291}--\blpage{2302}
(\byear{2011})
\end{barticle}
\endbibitem

\bibitem[\protect\citeauthoryear{Seghier et~al.}{2019}]{seghier2019advanced}
\begin{barticle}
\bauthor{\bsnm{Seghier}, \binits{A.}},
\bauthor{\bsnm{Li}, \binits{J.}},
\bauthor{\bsnm{Sun}, \binits{D.Z.}}:
\batitle{Advanced encryption standard based on key dependent s-box cube}.
\bjtitle{IET Information Security}
\bvolume{13}(\bissue{6}),
\bfpage{552}--\blpage{558}
(\byear{2019})
\end{barticle}
\endbibitem

\bibitem[\protect\citeauthoryear{Webster and Tavares}{1985}]{webster1985design}
\begin{botherref}
\oauthor{\bsnm{Webster}, \binits{A.}},
\oauthor{\bsnm{Tavares}, \binits{S.E.}}:
On the design of s-boxes.
Conference on the theory and application of cryptographic techniques,
523--534
(1985).
Springer
\end{botherref}
\endbibitem

\bibitem[\protect\citeauthoryear{Ao et~al.}{2017}]{ao2017construction}
\begin{barticle}
\bauthor{\bsnm{Ao}, \binits{T.}},
\bauthor{\bsnm{Rao}, \binits{J.}},
\bauthor{\bsnm{Dai}, \binits{K.}},
\bauthor{\bsnm{Zou}, \binits{X.}}:
\batitle{Construction of high quality key-dependent s-boxes}.
\bjtitle{Nonlinearity (Ns)}
\bvolume{13}(\bissue{14}),
\bfpage{15}
(\byear{2017})
\end{barticle}
\endbibitem

\bibitem[\protect\citeauthoryear{Sa{\u{g}}d{\i}{\c{c}}o{\u{g}}lu}{2003}]{saugdiccouglu2003cryptological}
\begin{botherref}
\oauthor{\bsnm{Sa{\u{g}}d{\i}{\c{c}}o{\u{g}}lu}, \binits{S.}}:
Cryptological viewpoint of boolean function.
Master's thesis,
Middle East Technical University
(2003)
\end{botherref}
\endbibitem

\bibitem[\protect\citeauthoryear{Canteaut et~al.}{2019}]{canteaut2019observations}
\begin{botherref}
\oauthor{\bsnm{Canteaut}, \binits{A.}},
\oauthor{\bsnm{K{\"o}lsch}, \binits{L.}},
\oauthor{\bsnm{Wiemer}, \binits{F.}}:
Observations on the dlct and absolute indicators.
Cryptology ePrint Archive
(2019)
\end{botherref}
\endbibitem

\bibitem[\protect\citeauthoryear{Dinur and Shamir}{2011}]{Dinur}
\begin{botherref}
\oauthor{\bsnm{Dinur}, \binits{I.}},
\oauthor{\bsnm{Shamir}, \binits{A.}}:
Breaking grain-128 with dynamic cube attacks.
Fast Software Encryption,
167--187
(2011)
\end{botherref}
\endbibitem

\bibitem[\protect\citeauthoryear{LI et~al.}{2023}]{li2023further}
\begin{botherref}
\oauthor{\bsnm{LI}, \binits{Z.}},
\oauthor{\bsnm{JIANG}, \binits{N.}},
\oauthor{\bsnm{ZHUO}, \binits{Z.}}:
Further results on autocorrelation of vectorial boolean functions.
IEICE Transactions on Fundamentals of Electronics, Communications and Computer Sciences,
2022--1096
(2023)
\end{botherref}
\endbibitem

\bibitem[\protect\citeauthoryear{Zhang and Zheng}{1996}]{zhang1996gac}
\begin{botherref}
\oauthor{\bsnm{Zhang}, \binits{X.-M.}},
\oauthor{\bsnm{Zheng}, \binits{Y.}}:
Gac—the criterion for global avalanche characteristics of cryptographic functions.
J. UCS The Journal of Universal Computer Science: Annual Print and CD-ROM Archive Edition Volume 1• 1995,
320--337
(1996)
\end{botherref}
\endbibitem

\bibitem[\protect\citeauthoryear{Matsui}{1993}]{matsui1993linear}
\begin{botherref}
\oauthor{\bsnm{Matsui}, \binits{M.}}:
Linear cryptanalysis method for des cipher.
Workshop on the Theory and Application of of Cryptographic Techniques,
386--397
(1993).
Springer
\end{botherref}
\endbibitem

\bibitem[\protect\citeauthoryear{Daemen and Rijmen}{2001}]{daemen2001wide}
\begin{botherref}
\oauthor{\bsnm{Daemen}, \binits{J.}},
\oauthor{\bsnm{Rijmen}, \binits{V.}}:
The wide trail design strategy.
Cryptography and Coding: 8th IMA International Conference Cirencester, UK, December 17--19, 2001 Proceedings 8,
222--238
(2001).
Springer
\end{botherref}
\endbibitem

\bibitem[\protect\citeauthoryear{Matsui}{1994}]{matsui1994first}
\begin{botherref}
\oauthor{\bsnm{Matsui}, \binits{M.}}:
The first experimental cryptanalysis of the data encryption standard.
Annual International Cryptology Conference,
1--11
(1994).
Springer
\end{botherref}
\endbibitem

\bibitem[\protect\citeauthoryear{Meier and Staffelbach}{1989}]{meier1989nonlinearity}
\begin{botherref}
\oauthor{\bsnm{Meier}, \binits{W.}},
\oauthor{\bsnm{Staffelbach}, \binits{O.}}:
Nonlinearity criteria for cryptographic functions.
Workshop on the Theory and Application of of Cryptographic Techniques,
549--562
(1989).
Springer
\end{botherref}
\endbibitem

\bibitem[\protect\citeauthoryear{Waqas et~al.}{2014}]{waqas2014generation}
\begin{botherref}
\oauthor{\bsnm{Waqas}, \binits{U.}},
\oauthor{\bsnm{Afzal}, \binits{S.}},
\oauthor{\bsnm{Mir}, \binits{M.A.}},
\oauthor{\bsnm{Yousaf}, \binits{M.}}:
Generation of aes-like s-boxes by replacing affine matrix.
2014 12th International Conference on Frontiers of Information Technology,
159--164
(2014).
IEEE
\end{botherref}
\endbibitem

\bibitem[\protect\citeauthoryear{Carlet et~al.}{2010}]{carlet2010vectorial}
\begin{botherref}
\oauthor{\bsnm{Carlet}, \binits{C.}},
\oauthor{\bsnm{Crama}, \binits{Y.}},
\oauthor{\bsnm{Hammer}, \binits{P.L.}}:
Vectorial boolean functions for cryptography.
(2010)
\end{botherref}
\endbibitem

\bibitem[\protect\citeauthoryear{Sarkar et~al.}{2019}]{sarkar2019relationship}
\begin{botherref}
\oauthor{\bsnm{Sarkar}, \binits{S.}},
\oauthor{\bsnm{Mandal}, \binits{K.}},
\oauthor{\bsnm{Saha}, \binits{D.}}:
On the relationship between resilient boolean functions and linear branch number of s-boxes.
International Conference on Cryptology in India,
361--374
(2019).
Springer
\end{botherref}
\endbibitem

\bibitem[\protect\citeauthoryear{Baksi et~al.}{2021}]{baksi2021default}
\begin{botherref}
\oauthor{\bsnm{Baksi}, \binits{A.}},
\oauthor{\bsnm{Bhasin}, \binits{S.}},
\oauthor{\bsnm{Breier}, \binits{J.}},
\oauthor{\bsnm{Khairallah}, \binits{M.}},
\oauthor{\bsnm{Peyrin}, \binits{T.}},
\oauthor{\bsnm{Sarkar}, \binits{S.}},
\oauthor{\bsnm{Sim}, \binits{S.M.}}:
Default: Cipher level resistance against differential fault attack.
Advances in Cryptology--ASIACRYPT 2021: 27th International Conference on the Theory and Application of Cryptology and Information Security, Singapore, December 6--10, 2021, Proceedings, Part II 27,
124--156
(2021).
Springer
\end{botherref}
\endbibitem

\bibitem[\protect\citeauthoryear{Dobbertin and Leander}{2004}]{dobbertin2004survey}
\begin{botherref}
\oauthor{\bsnm{Dobbertin}, \binits{H.}},
\oauthor{\bsnm{Leander}, \binits{G.}}:
A survey of some recent results on bent functions.
International Conference on Sequences and Their Applications,
1--29
(2004).
Springer
\end{botherref}
\endbibitem

\bibitem[\protect\citeauthoryear{Nizam~Chew and Ismail}{2020}]{2020}
\begin{barticle}
\bauthor{\bsnm{Nizam~Chew}, \binits{L.C.}},
\bauthor{\bsnm{Ismail}, \binits{E.S.}}:
\batitle{S-box construction based on linear fractional transformation and permutation function}.
\bjtitle{Symmetry}
\bvolume{12}(\bissue{5}),
\bfpage{826}
(\byear{2020})
\doiurl{10.3390/sym12050826}
\end{barticle}
\endbibitem

\bibitem[\protect\citeauthoryear{Mukherjee et~al.}{2021}]{mukherjee2021design}
\begin{bbook}
\bauthor{\bsnm{Mukherjee}, \binits{C.S.}},
\bauthor{\bsnm{Roy}, \binits{D.}},
\bauthor{\bsnm{Maitra}, \binits{S.}}:
\bbtitle{Design and Cryptanalysis of ZUC: A Stream Cipher in Mobile Telephony}.
\bpublisher{Springer}, \blocation{???}
(\byear{2021})
\end{bbook}
\endbibitem

\bibitem[\protect\citeauthoryear{Boura and Canteaut}{2018}]{Boura_Canteaut_2018}
\begin{barticle}
\bauthor{\bsnm{Boura}, \binits{C.}},
\bauthor{\bsnm{Canteaut}, \binits{A.}}:
\batitle{On the boomerang uniformity of cryptographic sboxes}.
\bjtitle{IACR Transactions on Symmetric Cryptology}
\bvolume{2018}(\bissue{3}),
\bfpage{290}--\blpage{310}
(\byear{2018})
\doiurl{10.13154/tosc.v2018.i3.290-310}
\end{barticle}
\endbibitem

\bibitem[\protect\citeauthoryear{Bao et~al.}{2019}]{bao2019peigen}
\begin{botherref}
\oauthor{\bsnm{Bao}, \binits{Z.}},
\oauthor{\bsnm{Guo}, \binits{J.}},
\oauthor{\bsnm{Ling}, \binits{S.}},
\oauthor{\bsnm{Sasaki}, \binits{Y.}}:
Peigen--a platform for evaluation, implementation, and generation of s-boxes.
IACR Transactions on Symmetric Cryptology,
330--394
(2019)
\end{botherref}
\endbibitem

\bibitem[\protect\citeauthoryear{Hasan et~al.}{2021}]{hasan2021c}
\begin{barticle}
\bauthor{\bsnm{Hasan}, \binits{S.U.}},
\bauthor{\bsnm{Pal}, \binits{M.}},
\bauthor{\bsnm{St{\u{a}}nic{\u{a}}}, \binits{P.}}:
\batitle{The c-differential uniformity and boomerang uniformity of two classes of permutation polynomials}.
\bjtitle{IEEE Transactions on Information Theory}
\bvolume{68}(\bissue{1}),
\bfpage{679}--\blpage{691}
(\byear{2021})
\end{barticle}
\endbibitem

\bibitem[\protect\citeauthoryear{Sarkar and Syed}{2018}]{sarkar2018bounds}
\begin{botherref}
\oauthor{\bsnm{Sarkar}, \binits{S.}},
\oauthor{\bsnm{Syed}, \binits{H.}}:
Bounds on differential and linear branch number of permutations.
Australasian conference on information security and privacy,
207--224
(2018).
Springer
\end{botherref}
\endbibitem

\bibitem[\protect\citeauthoryear{Preneel et~al.}{1991}]{preneel1991propagation}
\begin{botherref}
\oauthor{\bsnm{Preneel}, \binits{B.}},
\oauthor{\bsnm{Van~Leekwijck}, \binits{W.}},
\oauthor{\bsnm{Van~Linden}, \binits{L.}},
\oauthor{\bsnm{Govaerts}, \binits{R.}},
\oauthor{\bsnm{Vandewalle}, \binits{J.}}:
Propagation characteristics of boolean functions.
Advances in Cryptology—EUROCRYPT’90: Workshop on the Theory and Application of Cryptographic Techniques Aarhus, Denmark, May 21--24, 1990 Proceedings 9,
161--173
(1991).
Springer
\end{botherref}
\endbibitem

\bibitem[\protect\citeauthoryear{Makarim and Tezcan}{2014}]{makarim2014relating}
\begin{botherref}
\oauthor{\bsnm{Makarim}, \binits{R.H.}},
\oauthor{\bsnm{Tezcan}, \binits{C.}}:
Relating undisturbed bits to other properties of substitution boxes.
International workshop on lightweight cryptography for security and privacy,
109--125
(2014).
Springer
\end{botherref}
\endbibitem

\bibitem[\protect\citeauthoryear{Cid et~al.}{2018}]{cid2018boomerang}
\begin{botherref}
\oauthor{\bsnm{Cid}, \binits{C.}},
\oauthor{\bsnm{Huang}, \binits{T.}},
\oauthor{\bsnm{Peyrin}, \binits{T.}},
\oauthor{\bsnm{Sasaki}, \binits{Y.}},
\oauthor{\bsnm{Song}, \binits{L.}}:
Boomerang connectivity table: a new cryptanalysis tool.
Annual International Conference on the Theory and Applications of Cryptographic Techniques,
683--714
(2018).
Springer
\end{botherref}
\endbibitem

\bibitem[\protect\citeauthoryear{Langford and Hellman}{1994}]{langford1994differential}
\begin{botherref}
\oauthor{\bsnm{Langford}, \binits{S.K.}},
\oauthor{\bsnm{Hellman}, \binits{M.E.}}:
Differential-linear cryptanalysis.
Advances in Cryptology—CRYPTO’94: 14th Annual International Cryptology Conference Santa Barbara, California, USA August 21--25, 1994 Proceedings 14,
17--25
(1994).
Springer
\end{botherref}
\endbibitem

\bibitem[\protect\citeauthoryear{Jeong et~al.}{2022}]{JEONG2022101931}
\begin{barticle}
\bauthor{\bsnm{Jeong}, \binits{J.}},
\bauthor{\bsnm{Koo}, \binits{N.}},
\bauthor{\bsnm{Kwon}, \binits{S.}}:
\batitle{New differentially 4-uniform permutations from modifications of the inverse function}.
\bjtitle{Finite Fields and Their Applications}
\bvolume{77},
\bfpage{101931}
(\byear{2022})
\doiurl{10.1016/j.ffa.2021.101931}
\end{barticle}
\endbibitem

\bibitem[\protect\citeauthoryear{Jakobsen and Knudsen}{1997}]{jakobsen1997interpolation}
\begin{botherref}
\oauthor{\bsnm{Jakobsen}, \binits{T.}},
\oauthor{\bsnm{Knudsen}, \binits{L.R.}}:
The interpolation attack on block ciphers.
Fast Software Encryption: 4th International Workshop, FSE’97 Haifa, Israel, January 20--22 1997 Proceedings 4,
28--40
(1997).
Springer
\end{botherref}
\endbibitem

\bibitem[\protect\citeauthoryear{Calvi}{2005}]{calvi2005lectures}
\begin{botherref}
\oauthor{\bsnm{Calvi}, \binits{J.-P.}}:
Lectures on multivariate polynomial interpolation
(2005)
\end{botherref}
\endbibitem

\bibitem[\protect\citeauthoryear{Kocher et~al.}{2011}]{kocher2011introduction}
\begin{barticle}
\bauthor{\bsnm{Kocher}, \binits{P.}},
\bauthor{\bsnm{Jaffe}, \binits{J.}},
\bauthor{\bsnm{Jun}, \binits{B.}},
\bauthor{\bsnm{Rohatgi}, \binits{P.}}:
\batitle{Introduction to differential power analysis}.
\bjtitle{Journal of Cryptographic Engineering}
\bvolume{1},
\bfpage{5}--\blpage{27}
(\byear{2011})
\end{barticle}
\endbibitem

\bibitem[\protect\citeauthoryear{Singh et~al.}{2017}]{singh2017analysis}
\begin{barticle}
\bauthor{\bsnm{Singh}, \binits{A.}},
\bauthor{\bsnm{Agarwal}, \binits{P.}},
\bauthor{\bsnm{Chand}, \binits{M.}}:
\batitle{Analysis of development of dynamic s-box generation}.
\bjtitle{Comput. Sci. Inf. Technol}
\bvolume{5}(\bissue{5}),
\bfpage{154}--\blpage{163}
(\byear{2017})
\end{barticle}
\endbibitem

\bibitem[\protect\citeauthoryear{Tezcan}{2016}]{tezcan2016truncated}
\begin{botherref}
\oauthor{\bsnm{Tezcan}, \binits{C.}}:
Truncated, impossible, and improbable differential analysis of ascon.
Cryptology ePrint Archive
(2016)
\end{botherref}
\endbibitem

\bibitem[\protect\citeauthoryear{Naseer et~al.}{2024}]{naseer2024s}
\begin{botherref}
\oauthor{\bsnm{Naseer}, \binits{M.}},
\oauthor{\bsnm{Tariq}, \binits{S.}},
\oauthor{\bsnm{Riaz}, \binits{N.}},
\oauthor{\bsnm{Ahmed}, \binits{N.}},
\oauthor{\bsnm{Hussain}, \binits{M.}}:
S-box security analysis of nist lightweight cryptography candidates: A critical empirical study.
arXiv preprint arXiv:2404.06094
(2024)
\end{botherref}
\endbibitem

\end{thebibliography}
\end{document}